\documentclass[11pt]{article}
\usepackage{amssymb,amsmath,graphicx,epsfig,setspace,amsfonts,verbatim,natbib}

\usepackage{rotating,longtable,booktabs, colortbl}       % needed for rotating
\usepackage{mdwlist}
\usepackage[shortlabels]{enumitem}
\usepackage{lscape,afterpage}                    % supplies a landscape
\usepackage{appendix}
\usepackage{url,hyperref}                                            % for emails. Also, it links all your references...

\usepackage[standard]{ntheorem}
\usepackage[algo2e]{algorithm2e}

\theoremstyle{break}
\theorembodyfont{\normalfont}
\newtheorem{algorithm}[algocf]{Algorithm}

\begin{document}
	\title{\textsc{Flexible Bayesian Quantile Analysis of Residential Rental Rates}}
	\author{
		\hspace{0.3in} Ivan Jeliazkov \\ \hspace{0.3in} University of 
		California, Irvine \and
		\hspace{0.3in} Shubham Karnawat  \\ \hspace{0.2in} University of 
		California, Irvine 
		\and
		Mohammad Arshad Rahman \\ Indian Institute of Technology 
		Kanpur \and
		Angela Vossmeyer \\ Claremont McKenna College \\ and NBER\thanks{Email addresses: {\tt ivan@uci.edu}, {\tt skarnawa@uci.edu}, {\tt  marshad@iitk.ac.in},  and {\tt 
		angela.vossmeyer@cmc.edu}.} }
	\date{\today}
	\maketitle

\begin{abstract}
	\noindent This article develops a random effects quantile regression 
	model for panel data that allows for increased distributional 
	flexibility, multivariate heterogeneity, and time-invariant covariates in 
	situations where mean regression may be unsuitable. Our approach is 
	Bayesian and builds upon the generalized asymmetric Laplace distribution 
	to decouple the modeling of skewness from the quantile parameter. We 
	derive an efficient simulation-based estimation algorithm, demonstrate 
	its properties and performance in targeted simulation studies, and employ 
	it in the computation of marginal likelihoods to enable formal Bayesian 
	model comparisons. The methodology is applied in a study of U.S. 
	residential rental rates following the Global Financial Crisis. Our 
	empirical results provide interesting insights on the interaction between 
	rents and economic, demographic and policy variables, weigh in on key 
	modeling features, and overwhelmingly support the additional flexibility 
	at nearly all quantiles and across several sub-samples. The practical 
	differences that arise as a result of allowing for flexible modeling can 
	be nontrivial, especially for quantiles away from the median.

\vspace{0.5pc}
\noindent \texttt{Keywords:} Bayesian inference, generalized asymmetric Laplace distribution, Markov chain Monte Carlo, panel data, rental markets.
\end{abstract}

\newlength{\single}\setlength{\single}{1.0\baselineskip}\newlength{\double} %
\setlength{\double}{1.5\baselineskip}\baselineskip\double

%------------------------------------------------------------------------------
\section{Introduction}\label{sec:Intro}
%------------------------------------------------------------------------------

This paper aims to provide complementary methodological and empirical 
contributions to the quantile regression literature. On the methodological 
side, we develop a flexible Bayesian approach to random effects quantile 
regression based on a generalization of the asymmetric Laplace distribution, 
specify an efficient Markov chain Monte Carlo (MCMC) estimation algorithm, 
and present methods for formal model comparison of various nested and 
non-nested models that also enable us to assess the importance of flexible 
modeling. Our methods are readily motivated by the econometric challenges of 
studying U.S. residential rental rates and their dependence on unemployment 
and mortgage policies following the Global Financial Crisis. Our 
investigation deals with key features of the data, including considerable 
zip-code-level heterogeneity and skewness in the distribution of rents. The 
separation of modeling features into those that are practically relevant from 
those that are not warranted in this context is handled by model comparison.

\cite{Koenker-Basset-1978} introduced quantile regression as a minimization 
problem involving an asymmetrically weighted linear loss function, but 
subsequent work has noted the duality between that approach and modeling 
through a likelihood function built on the asymmetric Laplace (AL) 
distribution \citep{Koenker-Machado-1999, Yu-Moyeed-2001}. The latter 
approach becomes very potent when the AL distribution is expressed as a 
mixture of normal and exponential distributions 
\citep{Kozumi-Kobayashi-2011}. The mixture formulation permits estimation by 
simple, yet powerful, MCMC algorithms, and has enabled extensions of the 
quantile methodology to a variety of other settings including censored data 
\citep{Kozumi-Kobayashi-2011}, binary data \citep{Benoit-Poel-2010, 
Ojha-Rahman-2021}, ordinal outcomes \citep{Rahman-2016, Alhamzawi-2016, 
Maheshwari-Rahman-2023}, linear mixed models \citep{Luo-Lian-Tian-2012}, and 
panels of binary \citep{Rahman-Vossmeyer-2019, Bresson-etal-2021}, ordinal 
\citep{Alhamzawi-Ali-Longitudinal2018}, or dynamic censored data 
\citep{Kobayashi-Kozumi-2012}. Recent work by \cite{Goncalves-etal-2020} 
considered extensions to the case of dynamic quantile linear models. While 
the application of the AL distribution has unlocked a plethora of new 
research opportunities, the AL distribution itself is not without its 
limitations. For instance, the skewness is completely determined once a 
quantile is chosen and the mode of the distribution is always fixed at the 
value of the location parameter. These limitation can be circumvented by 
introducing a shape parameter into the mean of the normal kernel in the AL 
mixture representation leading to the generalized asymmetric Laplace (GAL) 
distribution \citep{Yan-Kottas-2017, Rahman-Karnawat-2019}.

We extend GAL modeling to the random effects panel setting by proposing an efficient MCMC sampler which offers a variety of algorithmic improvements through suitable transformations of the mixture variables, block sampling of scale and shape parameters, and block sampling of the individual-specific and common effect parameters \citep[cf.][]{Nascimento-Goncalves-2021}. These changes eliminate the problem of high autocorrelation in the MCMC draws, but are also applicable to the MCMC estimation models based on the simpler AL distribution \citep[cf.][]{Luo-Lian-Tian-2012} while also allowing for correlated random effects. For both the GAL and AL panel data models, we adapt the methods of \cite{Chib-1995} and \cite{Chib-Jeliazkov-2001} to enable model comparison through marginal likelihoods, which, with few exceptions \citep[e.g.,][]{Kobayashi-Kozumi-2012, Maheshwari-Rahman-2023} is broadly lacking in the quantile literature. Several simulation studies carefully examine the properties and practical appeal of the proposed techniques.

The empirical contribution of the paper involves the study of U.S. residential rental rates during the recovery period following the Global Financial Crisis. We construct a novel data set that includes median rental rates in $14,533$ zip codes from 2010 to 2016, as well as zip-code level  economic, demographic, mortgage, and tax policy controls. Our methodology is particularly appealing in this setting because housing prices and rents are heavily skewed and heterogeneous across regions. For instance, from 2010 to 2016, the Cleveland MSA region's change in ``All Transaction House Price Index'' was about 8.42, whereas the San Francisco MSA region's change was about 156.\footnote{Based on data from the FRED database at the Federal Reserve Bank of St.\! Louis.} Skewness, along with heterogeneity in economic outcomes, has been identified as an important driver of public policy and political economy considerations \citep{Benhabib-Bisin-2018}.

The data reveal a positive impact of unemployment on residential rental rates as uncertain job prospects reduce the willingness and ability of households to commit to homeownership and instead shift demand towards rental units. We also find that home mortgage deductions decrease rental prices by making homeownership more attractive. This finding is particularly relevant in the context of the Tax Cuts and Jobs Act, which was passed in 2017. The law lowered mortgage deductions, suggesting that one consequence of the policy change is an expected rise in rents. Lastly, model comparisons across many quantiles and samples reveal that the data overwhelmingly support the more flexible GAL modeling framework.

The remainder of the paper is organized as follows. In Section~\ref{sec:meth}, we present the proposed modeling, estimation and model comparison framework. This section also presents improved algorithms for the simpler AL-based model. Section~\ref{sec:simStudies} illustrates the proposed algorithms in multiple simulation studies. Section~\ref{sec:application} describes the data, presents our rental rates application and discusses the results, while Section~\ref{sec:conclusion} concludes.

%------------------------------------------------------------------------------
\section{Methodology}\label{sec:meth}
%------------------------------------------------------------------------------
This section introduces our proposed model, discusses the challenges 
associated with its estimation, and presents the MCMC estimation algorithm. 
The section also offers an improved algorithm for estimating AL-based models. 
Besides, this section describes the computation of marginal likelihood for 
the proposed framework and the AL-based model.

%------------------------------------------------------------------------------
\subsection{The Flexible Random Effects Quantile (FREQ) Model}\label{sec:FREQ}
%------------------------------------------------------------------------------

We focus on a panel data model which takes the form
%-------------------------------
\begin{equation}
y_{it} = x'_{it} \beta_{p_{0}} + z'_{it} \alpha_{i} + \varepsilon_{it}, \qquad
i = 1, \ldots, n,  \quad t  = 1, \ldots, T_{i},
\label{eq:FREQmodel}
\end{equation}
%-------------------------------
where $y_{it}$ denotes the $t$-th response on the $i$-th unit, $x_{it}$ is a 
$k$ vector of covariates, $\beta_{p_{0}}$ is a $k$ vector of common 
parameters at the $p_{0}$-th quantile (henceforth, simply $\beta$), $z_{it}$ 
is an $l$ vector of variables with $z_{it} \subseteq x_{it}$, and 
$\alpha_{i}$ is an $l$ vector of subject-specific random effects that induces 
dependence between observations on the same individual.\footnote{An 
unfortunate rift in terminology has persisted between statistics and 
econometrics in the panel (longitudinal) context. In  statistics, $\beta$ and 
$\{ \alpha_i \}$ are called fixed and random effects, respectively, because 
the former do not vary with $i$, whereas the latter are subject-specific. In 
econometrics these terms are used to distinguish between alternative ways of 
dealing with $\{ \alpha_i \}$ -- fixed effects estimators remove the 
heterogeneity (if possible) by data transformations such as mean- or 
first-differencing, whereas random effects estimators model the $\{ \alpha_i 
\}$ explicitly through a distribution.}

Although not immediately obvious from the notation, the setup is rather general and can capture dynamics, unknown covariate functions, and correlated random effects, depending on what is included in $x_{it}$ (and potentially also in $z_{it} \subseteq x_{it}$). In particular, dynamic modeling can be pursued by including lags of $y_{it}$ in $x_{it}$ and ensuring that the lag coefficients satisfy stationarity. Flexible functional modeling for some covariate $s$ can be implemented through a set of basis functions $\mathcal{B} = \{ b_1, \ldots, b_m \}$, e.g., B-splines, natural splines, truncated power series, wavelets, etc., \citep{Ruppert-etal-2003} so that $f(s) = \sum_{j=1}^{m} b_j(s) \delta_j$, in which case $x_{it}$ includes $(b_1 (s_i), \ldots, b_m (s_i) )'$, while $(\delta_1, \ldots, \delta_m )'$ becomes part of the regression parameter vector $\beta$. In addition, correlated random effect models where the heterogeneity can be correlated with certain observed covariates is handled by interacting those covariates with $z_{it}$ and including the result in $x_{it}$ \citep[see, e.g.,][]{Chamberlain-1984, Mundlak-1978, Chib-Jeliazkov-2006}. Random effect models are also indispensable in settings with multivariate heterogeneity or time-invariant covariates because the data transformations (i.e., mean- or first-differencing) underlying ``fixed effects'' estimators in econometrics (i) do not remove slope heterogeneity and (ii) wipe out any time-invariant covariates.

We parameterize the model in Equation~\eqref{eq:FREQmodel} by letting 
$\varepsilon_{it} \stackrel{iid}{\sim} \textrm{GAL}(0, \sigma, p_{0}, 
\gamma)$, using the quantile-fixed GAL distribution \citep{Yan-Kottas-2017, 
Rahman-Karnawat-2019} -- a generalization stemming from the mixture 
representation of the AL distribution -- to decouple the modeling of skewness 
from the quantile parameter. A variable $s$ is said to follow a 
quantile-fixed GAL distribution, i.e., $s \sim \textrm{GAL}(\mu, \sigma, 
p_{0}, \gamma)$, where $\mu$, $\sigma$, $p_{0}$, and $\gamma$ represent the 
location, scale, quantile, and skewness parameters, respectively, if it has 
density 
%-------------------------------
\begin{equation}
	\begin{split}
	f_{GAL} (s | \mu, \sigma, p_{0}, \gamma) &= \frac{2p(1-p)}{\sigma} \Bigg( \bigg[
	\Phi\left(- s^{\ast}\, \frac{p_{\gamma_{+}}}{|\gamma|} + \frac{p_{\gamma_{-}}}
	{p_{\gamma_{+}}} |\gamma|\right) -
	\Phi\left(\frac{p_{\gamma_{-}}}{p_{\gamma_{+}}}|\gamma| \right) \bigg] \\
	&\times \exp\bigg\{ - s^{\ast}\,  p_{\gamma_{-}} + 
	\frac{\gamma^2}{2}
	\bigg( \frac{p_{\gamma_{-}}}{p_{\gamma_{+}}} \bigg)^{2}
	\bigg\} I\left( \frac{s^{\ast}}{\gamma} > 0\right) \\
	&+ \Phi\left(- |\gamma| + s^{\ast}\, \frac{ p_{\gamma_{+}}}{|\gamma|}
	I\left( \frac{s^{\ast}}{\gamma} > 0\right) \right)
	\exp\bigg\{ - s^{\ast} \, p_{\gamma_{+}} + \frac{\gamma^2}{2} \bigg\}
	\Bigg), \label{eq:GALpdf2}
	\end{split}
\end{equation}
%-------------------------------
where $s^{\ast} = \frac{s - \mu}{\sigma}$, $p\equiv p(\gamma,p_{0})=I(\gamma<0) + [p_{0} - I(\gamma<0)]/g(\gamma)$, $p_{\gamma_{+}} = p - I(\gamma > 0)$ and $p_{\gamma_{-}} = p - I(\gamma < 0)$, $g(\gamma) = 2 \Phi(-|\gamma|) \exp(\gamma^{2}/2)$ and $\gamma \in (L,U)$, where $L$ is the negative square root of $g(\gamma)=1-p_{0}$ and $U$ is the positive square root of $g(\gamma)=p_{0}$ (see Section~2 in \citet{Rahman-Karnawat-2019} for more details). The term ``quantile-fixed'' refers to the fact that the density in Equation~\eqref{eq:GALpdf2} satisfies $\int_{-\infty}^{\mu} f_{GAL}( 
\varepsilon_{it} | \mu, \sigma, p_{0}, \gamma )d \varepsilon_{it} = p_{0}$. 
The special case of AL density results when 
$\gamma=0$, this is more clearly seen from the mixture representation 
presented in Equation~\eqref{eq:errorMixture}.

\begin{figure*}[tbh]
	\centerline{
		\mbox{\includegraphics[width=7.25in, height=2.4in]{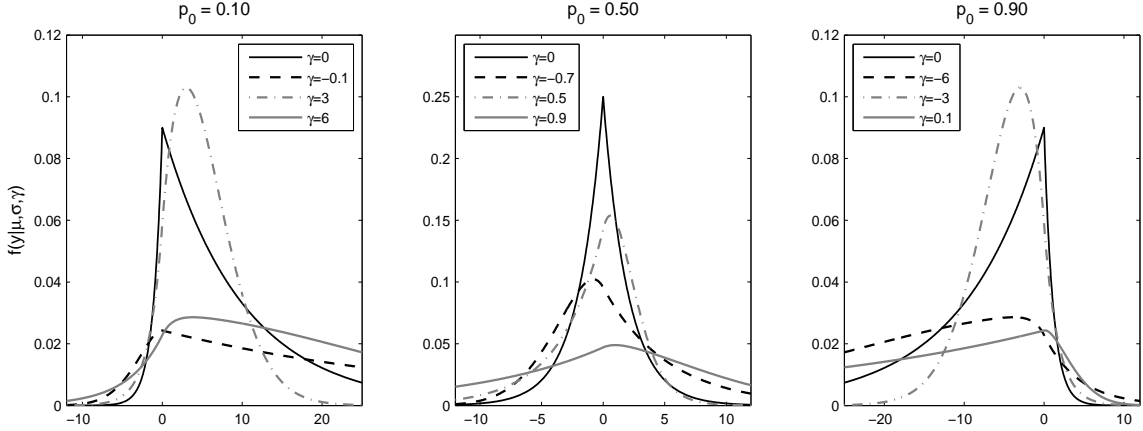}}
	}
	\caption{Probability density plots of the AL ($\gamma=0$) and GAL
		($\gamma \neq 0$) distributions.} \label{fig:galpdf}
\end{figure*}

Figure~\ref{fig:galpdf} offers a visualization of the differences between the 
quantile-fixed GAL and AL densities. The figure shows three different 
quantiles when $\sigma = 1$, the standard case. We observe that the GAL 
distribution, unlike the AL distribution, allows the mode to vary rather than 
being fixed at $\mu=0$ at all quantiles. Additionally, at the median 
$p_{0}=0.50$, the GAL distribution can be positively ($\gamma<0$) or 
negatively ($\gamma>0$) skewed and can have tails that are heavier or 
narrower than the AL distribution. These characteristics make GAL 
significantly more flexible than AL, but the value of the extra flexibility 
is an application-specific empirical question.
%---------------------------  Figure 1 ---------------------------------------
%-------------------------------------------------------------------------------

Because of the additional flexibility that the GAL distribution offers over the AL distribution, we refer to the model based on the former as the Flexible Random Effects Quantile (FREQ) model, and the model based on the latter as the Random Effects Quantile (REQ) model.

The distributional assumption on the error term, $\varepsilon_{it} 
\stackrel{iid}{\sim} 
\textrm{GAL}(0,\sigma,p_{0},\gamma)$ implies that $y_{it}|\alpha_{i} 
\stackrel{ind}{\sim} \textrm{GAL}(x'_{it} \beta + z'_{it} 
\alpha_{i},\sigma,p_{0},\gamma)$ for $i = 1, \ldots, n$ and $t = 1, \ldots, 
T_{i}$. Assuming a density $f(\alpha|\Omega)$ for the random effects, 
the complete data likelihood can be expressed as
%------------------------------
\begin{equation}
f(y,\alpha|\beta,\sigma, \gamma, \varphi^{2}) = \prod_{i=1}^{n} \bigg[
\bigg\{ \prod_{j=1}^{T_{i}} f_{GAL} \left( y_{it} | x'_{it} \beta + z'_{it} \alpha_{i},
\sigma, p_{0}, \gamma \right) \bigg\} f(\alpha_{i}|\Omega) \bigg],
\label{eq:fulllikelihood}
\end{equation}
%------------------------------
where $\alpha = (\alpha_{1}, \ldots, \alpha_{n})$, $y = (y_{1}, \ldots, 
y_{n})$ with each $y_{i} = (y_{i1}, \ldots, y_{iT_{i}})'$ for $i = 1,\ldots, 
n$. The density $f(\alpha_{i}|\Omega)$ can be any suitable distribution, 
but is typically assumed normal \citep{Luo-Lian-Tian-2012}, e.g., here we let 
$\alpha_{i}|\Omega \stackrel{iid}{\sim} N(0_{l}, \Omega)$ for 
$i = 1, \ldots, n$. The complete data likelihood in 
Equation~\eqref{eq:fulllikelihood} can be combined with priors on the 
parameters to obtain the joint posterior distribution, but this posterior 
does not yield known conditional posteriors suitable for a tractable MCMC 
algorithm. Hence, we utilize the mixture representation of the GAL 
distribution obtained by introducing a shape parameter into the mean of the 
normal kernel in the normal-exponential mixture of the AL distribution and 
mixing with respect to a half-normal distribution \citep{Yan-Kottas-2017, 
Rahman-Karnawat-2019}.

For $\varepsilon_{it} \sim \textrm{GAL}(0,\sigma,p_{0},\gamma)$, the mixture representation can be expressed as,
%------------------------------
\begin{equation}
\varepsilon_{it} =  \sigma A \omega_{it} + \sigma C |\gamma|s_{it} +
\sigma (B \omega_{it})^{\frac{1}{2}} u_{it},
\label{eq:errorMixture}
\end{equation}
%------------------------------
where $s_{it} \sim N^{+}(0,1)$, $\omega_{it} \sim \mathcal{E}(1)$, $u_{it}
\sim N(0,1)$, $A\equiv A(p)=\frac{1-2p}{p(1-p)}$, $B\equiv
B(p)=\frac{2}{p(1-p)}$, $C=[I(\gamma>0)-p]^{-1}$, and $p$ is as defined 
earlier. Here, $N^{+}, \mathcal{E}, N$ denote half-normal,
exponential, and normal distributions, respectively. Note that the GAL 
mixture distribution reduces to an AL mixture distribution when $\gamma$ is 
set to 0, as mentioned earlier. Substituting the mixture
representation given by Equation~\eqref{eq:errorMixture} into 
Equation~\eqref{eq:FREQmodel}, the model can be written as 
%------------------------------
$y_{it} = x'_{it}\beta + z'_{it}\alpha_{i} +
\sigma A\omega_{it} +  \sigma C |\gamma|s_{it} + \sigma (B
\omega_{it})^{\frac{1}{2}}u_{it}.
$
%------------------------------
In this formulation, the scale parameter appears in the conditional mean which is not suitable for estimation \citep{Kozumi-Kobayashi-2011}. Therefore, we make the following transformation $h_{it} = \sigma s_{it}$, $\nu_{it} =
\sigma \omega_{it}$ and rewrite the model as,
%------------------------------
\begin{equation}
y_{it} = x'_{it}\beta + z'_{it}\alpha_{i} +   A \nu_{it} + C |\gamma|h_{it} +
(\sigma B \nu_{it})^{\frac{1}{2}}u_{it}.
\label{eq:mixtureModel2}
\end{equation}
%------------------------------
Stacking the model given by Equation~\eqref{eq:mixtureModel2} for each 
individual $i$, we get
%------------------------------
\begin{equation}
y_{i} = X_{i}\beta + Z_{i}\alpha_{i} + A\nu_{i} + C|\gamma|h_{i} +
\Lambda_{i}^{1/2} u_{i},
\label{eq:stackedModel}
\end{equation}
%------------------------------
where $y_{i} = (y_{i1}, \cdots, y_{iT_{i}})'$,  $ X_{i} = ( x'_{i1}, \cdots, 
x'_{i T_{i}} )'$ is the design matrix of size $ T_{i} \times k $ for each 
individual $i$, $Z_{i} = (z'_{i1}, \cdots, z'_{iT_{i}})'$ is $T_{i} \times l$ 
matrix of covariates associated with the random effects, $\nu_{i} = 
(\nu_{i1}, \cdots, \nu_{i T_{i} } )'$, $ h_{i} = (h_{i1}, \cdots, 
h_{iT_{i}})'$, $u_{i} = (u_{i1}, \cdots, u_{i T_{i}})'$  and the diagonal 
matrix
%------------------------------
\begin{displaymath}
\Lambda_{i} =
\begin{bmatrix}
\sigma B \nu_{i1}  & 0 & \cdots&  0 \\
0 & \sigma B \nu_{i2} & \cdots & 0 \\
\vdots & \vdots & \ddots & \vdots \\
0& 0& \cdots & \sigma B \nu_{i T_{i}}
\end{bmatrix}.
\end{displaymath}
%------------------------------
The model in Equation~\eqref{eq:stackedModel} implies that 
$y_{i}|\beta,\alpha_{i}, \nu_{i}, h_{i}, \sigma, \gamma \sim N_{T_{i} } 
\left( X_{i}\beta + Z_{i}\alpha_{i} + A\nu_{i} + C|\gamma|h_{i}, 
\hspace{.05in} \Lambda_{i}	\right)$, which is combined with the priors
%------------------------------
\begin{equation}
%\begin{split}
\beta \sim N(\beta_{0}, B_{0}), \hspace{.25in}
\sigma \sim IG\left( \frac{n_{0}}{2}, \frac{d_{0}}{2} \right),
\hspace{0.25in} \gamma \sim Unif(L,U), \hspace{0.25in}
\Omega \sim IW(\omega_{0}, O_{0})
%\varphi^2 \sim IG\left( \frac{c_{1}}{2}, \frac{d_{1}}{2} \right),
%\end{split}
\label{eq:priors}
\end{equation}
%------------------------------
to obtain the joint posterior distribution. In the notation above, $IG$ and 
$IW$ denote an inverse-Gamma and inverse-Wishart distribution, respectively. 
Let $\Theta=(\beta,\alpha,\nu,h,\sigma, \gamma, \Omega)$, then the 
posterior distribution can be expressed as follows,
%------------------------------
\begin{align}
& \pi(\Theta|y) \propto f(y|\Theta) \times
 \pi(\alpha|\Omega) \times \pi(\nu) \times \pi(h) \times \pi(\beta) \times
\pi(\sigma) \times \pi(\Omega) \times \pi(\gamma)  \notag \\
%----------
& \propto \prod_{i=1}^{n} \Bigg[ |\Lambda_{i}|^{-1/2} \exp \bigg\{ -\frac{1}{2}
\Big( y_{i} - X_{i}\beta -Z_{i}\alpha_{i} - A\nu_{i} - C|\gamma|h_{i} \Big)'
\Lambda_{i}^{-1} \Big( y_{i} - X_{i}\beta -Z_{i}\alpha_{i}  - A\nu_{i}
\notag \\
%----------
&	\hspace{.2in}  -  C|\gamma|h_{i} \Big)	\bigg\}	
\times |\Omega|^{-1/2}\exp{ \bigg\{-\frac{1}{2} \;
\alpha'_{i}\Omega^{-1}\alpha_{i} \bigg\} }
\times  \bigg( \sigma^{-T_{i}} \exp{
\Big\{- \sum_{j=1}^{T_{i}} \frac{ \nu_{it}}{\sigma} \Big\} } \bigg)
\notag \\
%----------
& 	\hspace{.2in}    \times \bigg( \sigma^{-T_{i}} \exp \left\{
-\frac{h'_{i}h_{i}}{2 \sigma^2}	 \right\} \bigg) \Bigg] \times
\exp \bigg\{-\frac{1}{2} (\beta -\beta_{0})' B_{0}^{-1} (\beta - \beta_{0})	
\bigg\}  \notag \\
%----------
& \hspace{.2in} \times  \left( \sigma^{- \frac{n_{0}}{2}-1} \exp\left\{
-\frac{d_{0}}{2 \sigma}	\right\}	\right) \times
|\Omega|^{-(\omega_{0}+l+1)/2} \exp \left\{ -\frac{1}{2} \, 
\textrm{tr}(\Omega^{-1} O_{0}) \right\}   ,
\label{eq:jointPosterior}
\end{align}
%------------------------------
where $f(y|\Theta)$ denotes the density, conditional on $\alpha$, resulting
from the stacked FREQ model given by Equation~\eqref{eq:stackedModel}.

The FREQ model has several appealing properties: (i) it can accommodate both common and random effect parameters, (ii) random effects can be associated with multiple variables, in addition to the constant, allowing for both slope and intercept heterogeneity, (iii) quantile regression gives us the ability to explore the entire distribution of the outcome variable $y$, and (iv) flexibility in the skewness parameter allows better fit for various settings. Economic data that are skewed, exhibit power laws, or present odd asymmetries would benefit from the model offered in this paper. Such data include distributions of income, bank assets, social networks, and house or rental prices, the latter of which is explored in this paper. 

%------------------------------------------------------------------------------
\subsection{Estimation}\label{sec:Estimation}
%------------------------------------------------------------------------------

The FREQ model can be estimated by sampling the objects of interest,
$(\beta,\alpha,\nu,h,\sigma,\gamma,\Omega)$ from their respective
conditional posterior densities as in Algorithm~\ref{alg:algorithm1}. We first sample the parameters 
$(\beta,\alpha)$ in a block conditional on 
remaining parameters where $\beta$ is sampled marginally of $\alpha$, and 
then $\alpha$ is sampled conditional on $\beta$. Both the conditional 
posteriors follow a normal distribution with
hyperparameters updated as shown in Algorithm~\ref{alg:algorithm1}. We 
utilize block sampling for two reasons: (i) to account for possible 
correlation between the two parameters, and (ii) to reduce the inefficiency 
factors in the MCMC draws \citep{Chib-Carlin-1999,Greenberg-2012}. 

%------------------------------------------------------------------------------
\begin{table*}[h!]
\begin{algorithm}[Sampling in the FREQ model] \label{alg:algorithm1}
\rule{\textwidth}{0.5pt} \small{
\begin{enumerate}[(1)]
%-------------------------------------------------
\item Sample $\beta $ and $\alpha$ in a block because of possible correlation between
the parameters.
\begin{enumerate}[(a)]
\item[(a)] Let $V_{i} = Z_{i} \Omega Z'_{i} + \Lambda_{i}$. Sample $\beta$
marginally of $\alpha$ as $\beta|y,\nu,h,\sigma,\gamma, \Omega \sim
N( \tilde{\beta},\tilde{B})$, where,
%--------------------------
\begin{equation*}
\tilde{B}^{-1} = \bigg(\sum_{i=1}^{n} X'_{i}V_{i}^{-1}X_{i} + B_{0}^{-1} \bigg)
\hspace{.1in} \mathrm{and} \hspace{.1in}
\tilde{\beta} = \tilde{B}\bigg( \sum_{i=1}^{n} X'_{i}V_{i}^{-1}(y_{i}- A\nu_{i}
- C|\gamma| h_{i})  + B_{0}^{-1} \beta_{0} \bigg).
\end{equation*}
%--------------------------
%-------------------------------------------------				
\item[(b)] Sample $\alpha_{i}|y, \beta, \nu, h, \sigma, \gamma, \Omega$
$\sim N( \tilde{\alpha_{i}}, \tilde{A_{i}} )$, where,
%---------------------------
\begin{equation*}
\tilde{A_{i}}^{-1} = \left( Z'_{i}\Lambda_{i}^{-1}Z_{i} + \Omega^{-1} \right)
\hspace{0.15in} \mathrm{and} \hspace{0.15in}
\tilde{a}_{i} = \tilde{A_{i}} \left( Z'_{i}\Lambda_{i}^{-1} (y_{i} -
X_{i}\beta - A\nu_{i} - C|\gamma|h_{i} ) \right).
\end{equation*}
%---------------------------	
\end{enumerate}
\item Sample $\Omega|\alpha,y$ from $IW (\tilde{\omega}, \tilde{O})$,
where, $\tilde{\omega} = n + \omega_{0}
\hspace{0.05in} \mathrm{and} \hspace{0.05in}
\tilde{O} = \sum_{i=1}^{n} \alpha_{i}\alpha'_{i} + O_{0} $.
%--------------------------	

%-------------------------------------------------
\item Jointly sample $(\sigma,\gamma)$ marginally of ($\nu, h$) using
a random-walk MH algorithm. The proposed draw $(\sigma',\gamma')$ is generated
from a bivariate truncated normal distribution $BTN_{(0, \infty) \times (L,U)}
\big((\sigma_{c}, \gamma_{c}), \iota^{2} \hat{D} \big)$, where $(\sigma_{c},
\gamma_{c})$ are the current values, $\iota$ is the tuning factor and $\hat{D}$
is the negative inverse of the Hessian obtained by maximizing the 
log-likelihood in Equation~\eqref{eq:fulllikelihood} with respect
to $(\sigma,\gamma)$ with $\beta$ set at the pooled ordinary least squares estimate.
We accept $(\sigma',\gamma')$ with MH probability,
%--------------------------
\begin{equation*}
\alpha_{MH}(\sigma_{c},\gamma_{c}; \sigma',\gamma') = \min \bigg\{0,\ln\bigg[
\frac{f(y, \alpha|\beta,  \sigma', \gamma') \, \pi(\beta,
\sigma', \gamma')} {f(y, \alpha|\beta, \sigma_{c},\gamma_{c}) \,
\pi(\beta, \sigma_{c},\gamma_{c} )} \;
\frac{\pi(\sigma_{c}, \gamma_{c} | (\sigma',\gamma'), \iota^{2} \hat{D})}
{\pi(\sigma',\gamma'| (\sigma_{c}, \gamma_{c}),\iota^{2} \hat{D} )}\bigg] \bigg\};
\end{equation*}
%--------------------------
else, repeat the current value $(\sigma_{c},\gamma_{c})$
in the next MCMC iteration. In the equation above, $f(y,\alpha|\cdot)$ 
denotes the full likelihood given in Equation~\eqref{eq:fulllikelihood}, 
$\pi(\beta, \sigma,\gamma )$ denotes the prior distributions given by 
Equation~\eqref{eq:priors}, and $\pi(\sigma_{c}, \gamma_{c} |
(\sigma',\gamma'), \iota^{2} \hat{D})$ denotes a bivariate truncated normal
probability with mean $(\sigma',\gamma')$ and covariance $\iota^{2} \hat{D}$. 
The quantity $\pi(\sigma',\gamma'| (\sigma_{c}, \gamma_{c}),\iota^{2} \hat{D} 
)$ has a similar interpretation.
%-------------------------------------------------
%-------------------------------------------------
%-------------------------------------------------						
\item Sample $\nu_{it}|y_{it}, \beta, \alpha_{i}, h_{it}, \sigma, \gamma $ $
    \sim GIG (\frac{1}{2}, \chi,  \psi_{\nu_{it}} )$ for all values of $i$
    and $t$, where,
%--------------------------
\begin{equation*}
\chi = \bigg(\frac{A^{2}}{\sigma B} + \frac{2}{\sigma} \bigg)
\hspace{0.25in} \mathrm{and} \hspace{0.25in}
\psi_{\nu_{it}} = \frac{(y_{it}-x'_{it}\beta - z'_{it}\alpha_{i} -
	C|\gamma|h_{it})^{2}}{\sigma B}.
\end{equation*}
%-----------------------------------------------------------------------------
\item Sample $h_{it}|y_{it}, \beta, \nu_{it}, \sigma, \gamma
\sim N^{+} ( \mu_{h_{it}}, \sigma^2_{h_{it}} )$ for all values of $i$ and 
$t$, where,
%--------------------------
\begin{equation*}
\sigma_{h_{it}}^{2} = \bigg(\frac{1}{\sigma^2} + \frac{C^{2}
\gamma^{2}}{\sigma  B \nu_{it}}  \bigg)^{-1}
\hspace{0.25in} \mathrm{and} \hspace{0.25in}
\mu_{h_{it}} = \sigma_{h_{it}}^{2} \bigg( \frac{ C|\gamma|
(y_{it}- x'_{it}\beta - z'_{it}\alpha_{i} -A\nu_{it})}{\sigma B \nu_{it}}  \bigg).
\end{equation*}
%--------------------------
\end{enumerate}
\rule{\textwidth}{0.5pt} }
\end{algorithm}
\end{table*}
%------------------------------------------------------------------------------

The random effects covariance matrix, $\Omega$, is sampled from an 
inverse-Wishart distribution with updated hyperparameters. The scale and 
shape parameters $(\sigma,\gamma)$ are jointly sampled marginally of $(\nu, 
h)$ using a random-walk Metropolis-Hastings (MH) algorithm 
\citep{Chib-Greenberg-1995}. Here, the target density is the product of the 
GAL likelihood and the prior distributions, given by 
Equation~\eqref{eq:fulllikelihood} and Equation~\eqref{eq:priors}, 
respectively; while the proposal values are drawn from a bivariate
truncated normal distribution. We note that joint sampling of 
$(\sigma,\gamma)$ is critical in reducing the autocorrelation of MCMC draws 
and hence to the efficiency of the algorithm. The mixture variable, $\nu$, is 
sampled from a generalized inverse Gaussian (GIG) distribution, draws from 
which are generated using the technique in \citet{Devroye-2014}. Lastly, the 
mixture variable, $h$ conditional on remaining 
parameters, is sampled from a half-normal distribution with updated 
hyperparameters. The derivations of the conditional posterior distributions 
and further details of Algorithm~\ref{alg:algorithm1} are presented in 
\ref{AppendixA:FREQ}

 %Algorithm for REQ
%------------------------------------------------------------------------------
\begin{table*}[!h]
\begin{algorithm}[Sampling in the REQ model] \label{alg:algorithm2}
\rule{\textwidth}{0.5pt} \small{
\begin{enumerate}[(1)]
%-------------------------------------------------
\item Sample $\beta $ and $\alpha$ in a block to account for possible correlation
between them.

\begin{enumerate}[(a)]
\item[(a)] Sample $\beta $ marginally of $\alpha$ as $\beta$
$|y, \nu,\sigma, \Omega \sim N( \tilde{\beta}, \tilde{B} ) $,
where,
%--------------------------
\begin{equation*}
\tilde{B}^{-1} = \bigg(\sum_{i=1}^{n} X'_{i}V_{i}^{-1}X_{i} +
B_{0}^{-1} \bigg)
\hspace{.1in} \mathrm{and} \hspace{.1in}
\tilde{\beta} = \tilde{B}\bigg( \sum_{i=1}^{n} X'_{i}V_{i}^{-1}(y_{i}-
A\nu_{i} )+ B_{0}^{-1} \beta_{0} \bigg).
\end{equation*}
%--------------------------
%-------------------------------------------------				
\item[(b)] Sample $\alpha_{i}|y, \beta, \nu, \sigma, \Omega$
$\sim N( \tilde{\alpha_{i}}, \tilde{A_{i}} )$, where,
%---------------------------
\begin{equation*}
\tilde{A_{i}}^{-1} = \left( Z'_{i}\Lambda_{i}^{-1}Z_{i} + \Omega^{-1}
\right)
\hspace{0.15in} \mathrm{and} \hspace{0.15in}
\tilde{a}_{i} = \tilde{A_{i}} \left( Z'_{i}\Lambda_{i}^{-1} (y_{i} -
X_{i}\beta - A\nu_{i} ) \right).
\end{equation*}
%---------------------------	
\end{enumerate}
\item Sample $\Omega|\alpha$ as $IW
(\tilde{\omega}, \tilde{O})$,
where, $ \tilde{\omega} = n + \omega_{0}
\hspace{0.05in} \mathrm{and} \hspace{0.05in}
\tilde{O} = \sum_{i=1}^{n} \left( \alpha_{i}\alpha'_{i} \right) + O_{0}.$
%--------------------------	
%-------------------------------------------------						
\item Sample $\nu_{it}$ conditional on model parameters as
$\nu_{it}|y_{it}, \beta, \alpha_{i}, h_{it}, \sigma, \gamma $
$ \sim GIG (\frac{1}{2}, \chi,  \psi_{\nu_{it}} ) $,
%--------------------------
\begin{equation*}
\chi = \bigg(\frac{A^{2}}{\sigma B} + \frac{2}{\sigma} \bigg)
\hspace{0.25in} \mathrm{and} \hspace{0.25in}
\psi_{\nu_{it}} = \frac{ (y_{it}-x'_{it}\beta - z'_{it}\alpha_{i}) }{\sigma B}.
\end{equation*}
%--------------------------
%-------------------------------------------------
\item Sample the scale parameter  $\sigma|y, \beta, \alpha, \nu \sim
IG\left( \tilde{n}/2, \tilde{d}/2 \right) $, where,
%--------------------------
\begin{equation*}
\tilde{n} = 3 \sum_{i=1}^{n} T_{i} + n_{0}
\hspace{0.25in} \mathrm{and} \hspace{0.25in}
\tilde{d} = \sum_{i=1}^{n}\sum_{t=1}^{T_{i}} \left\{ \frac{\left(	
y_{it}-x'_{it}\beta-z'_{it}\alpha_{i} - A\nu_{it}	\right)^2}{B\nu_{it}} +2 \nu_{it}
\right\} + d_{0}.
\end{equation*}
%--------------------------
\end{enumerate}
\rule{\textwidth}{0.5pt} }
\end{algorithm}
\end{table*}
%------------------------------------------------------------------------------

To exemplify the practical utility of the FREQ model, we also estimate the 
more established REQ model using the sampler presented in 
Algorithm~\ref{alg:algorithm2}. This sampling algorithm has two important 
improvements from the sampler proposed in \citet{Luo-Lian-Tian-2012}. First, 
$(\beta, \alpha)$ are sampled in a single block which significantly lowers 
the autocorrelation in the MCMC draws and improves the mixing of the Markov 
chain. Because we can achieve lower inefficiency factors in 
Algorithm~\ref{alg:algorithm2}, the number of MCMC draws can typically be 
reduced, thereby decreasing computational burdens and run times. Second, we 
correct the updating for $\sigma$ by including the terms involving the 
exponential variable in the updated hyperparameters. The resulting MCMC 
algorithm is fast, efficient, and maintains the tractability of the sampling 
distributions.

Before ending this section, we note that the FREQ model presented in 
Section~\ref{sec:FREQ} and the estimation procedure explained above (and that 
of REQ model) considers multivariate heterogeneity i.e., random effects in 
the intercept and slope parameters. This leads to the distribution, 
$\alpha_{i} \sim N(0_{l}, \Omega)$, where $\Omega$ is an $l \times l$ 
covariance matrix that is assigned an inverse-Wishart prior distribution. In 
the special case of intercept heterogeneity only, we have $\alpha_{i} \sim 
N(0, \varphi^{2})$ for $i=1,2, \cdots,n$. Since $\varphi^{2}$ is a scalar, it 
is assigned an $IG$ prior distribution i.e., $\varphi^{2} \sim 
IG(c_{1}/2, d_{1}/2)$. This leads to changes in the conditional 
posterior density of $(\beta, \alpha)$ and the conditional posterior of 
$\Omega$ is replaced by that of $\varphi^{2}$. Specifically, $\beta$ is 
sampled as in Step~1(a) of Algorithm~\ref{alg:algorithm1} except that $V_{i} 
= \varphi^{2} Z_{i}Z'_{i} + \Lambda_{i}$; whereas for Step~1(b), $\alpha_{i}$ 
is sampled from $N( \tilde{\alpha_{i}}, \tilde{A_{i}} )$, 
where $\tilde{A_{i}}^{-1} = \left( Z'_{i}\Lambda_{i}^{-1}Z_{i} + 
(\varphi^{2})^{-1} I_{l} \right)$ and the expression for $\tilde{a}_{i}$ is 
unaltered. The parameter $\varphi^2$ is sampled from an updated $IG 
(\tilde{c}_{1}/2,\tilde{d}_{1}/2)$, where, $\tilde{c}_{1} = nl+ c_{1}$ and
$\tilde{d}_{1} = \sum_{i=1}^{n} \alpha'_{i}\alpha_{i} + d_{1} $. Step~(3) of 
Algorithm~\ref{alg:algorithm1} remains largely unaltered, 
except that in the full likelihood given by  
Equation~\eqref{eq:fulllikelihood}, the density $f(\alpha_{i}|\Omega)$ is 
replaced by $f(\alpha_{i}|\varphi^{2})$. Finally, Step~(4) and Step~(5) of 
Algorithm~\ref{alg:algorithm1} remains unchanged, except that $z_{it} = 1$ 
and $\alpha_{i}$ is a scalar. This modified MCMC algorithm 
is utilized to estimate the residential rental rates model in 
Section~\ref{sec:application}, where intercept heterogeneity is of 
considerable importance.

Similarly, estimating the REQ model with intercept heterogeneity
requires some modification to Algorithm~\ref{alg:algorithm2}. In particular, 
the sampling of $\beta$ in Step~1(a) now requires $V_{i} 
= \varphi^{2} Z_{i}Z'_{i} + \Lambda_{i}$; while sampling $\alpha_{i}$ in 
Step~1(b) needs $\tilde{A_{i}}^{-1} = \left( Z'_{i}\Lambda_{i}^{-1}Z_{i} + 
(\varphi^{2})^{-1} I_{l} \right)$. All other expression in the conditional 
posteriors remain unchanged. The parameter $\varphi^2|\alpha,y$ is sampled 
from an $IG(\tilde{c}_{1}/2, \tilde{d}_{1}/2)$,
where $ \tilde{c}_{1} = nl+ c_{1}$ and $\tilde{d}_{1} = \sum_{i=1}^{n} 
\left(	\alpha'_{i}\alpha_{i} \right) + d_{1}$. Finally, Step~(3) and 
Step~(4) of Algorithm~\ref{alg:algorithm2} remains unchanged, except note 
that $z_{it} = 1$ and $\alpha_{i}$ is a scalar. This modified algorithm for 
the REQ model is also used to estimate the rental rates application in 
Section~\ref{sec:application} for comparison.

%------------------------------------------------------------------------------
\subsection{Bayesian Model Comparison and Marginal Likelihood Estimation}\label{sec:ml}
%------------------------------------------------------------------------------

To properly address model uncertainty, Bayesian model comparison proceeds by 
representing the posterior model probability of model $\mathcal{M}_{s}$ given 
the data $y$ as,
\[
\Pr(\mathcal{M}_{s}| y ) \propto \Pr(\mathcal{M}_{s}) m(y|\mathcal{M}_{s}),
\]
where $\Pr(\mathcal{M}_{s})$ is the prior model probability and 
$m(y|\mathcal{M}_{s})$ is the marginal likelihood. Given the sampling density 
$f(y|\mathcal{M}_{s}, \Theta_{s})$ and prior distribution 
$\pi(\Theta_{s}|\mathcal{M}_{s})$ under model $\mathcal{M}_{s}$, the marginal 
likelihood is defined as the integral 
\[
m(y|\mathcal{M}_{s}) = \int f(y|\mathcal{M}_{s},\Theta_{s}) \pi(\Theta_{s}|\mathcal{M}_{s}) \, d\Theta_{s},
\]
which can also be expressed, using Bayes' theorem, as
%-----------------------------
\begin{equation}
m(y|\mathcal{M}_{s}) = \frac{f(y|\mathcal{M}_{s},\Theta_{s}) \,
\pi(\Theta_{s}|\mathcal{M}_{s})}{\pi(\Theta_{s} | y, \mathcal{M}_{s})},
\label{eq:BMLI}
\end{equation}
%-----------------------------
where the numerator is the product of the likelihood function and prior
density, and the denominator is the joint posterior density \citep{Chib-1995, Chib-Jeliazkov-2001}. Equation~\eqref{eq:BMLI} is known as the \emph{basic marginal likelihood identity} since it holds for all values in the parameter space. However, marginal likelihood estimate is typically computed at a high-density point (such as the mean or mode), denoted $\Theta^{\ast}_{s}$, to minimize estimation variability. The numerator quantities in Equation~\eqref{eq:BMLI} are generally directly available, and therefore, the problem of marginal likelihood estimation is reduced to finding an estimate of the posterior ordinate in the denominator of Equation~\eqref{eq:BMLI}.

Well-known properties of Bayesian model comparisons based on marginal 
likelihoods and their ratios, or Bayes factors, are that they lead to 
finite-sample model probabilities, do not require competing models to be 
nested and have appealing asymptotic properties that give rise to information 
criteria \citep{Greenberg-2012}. Another important, yet underappreciated, 
point is that marginal likelihoods provide a measure of sequential out of 
sample predictive fit, which can be seen by writing
%-------------------------------
\begin{eqnarray*}
	m(y|\mathcal{M}_{s}) &=& \prod_{i=1}^{n} m(y_i|  \{ y_j \}_{j<i}, \mathcal{M}_{s}) \\
	&=& \prod_{i=1}^{n} \int f(y_i | \{ y_j \}_{j<i}, \Theta_{s}, 
	\mathcal{M}_{s}) \pi(\Theta_{s} | \{ y_j \}_{j<i}, \mathcal{M}_{s}) \, d 
	\Theta_{s}.
\end{eqnarray*}
%-------------------------------
Therefore, the adequacy of the model as captured by the marginal likelihood 
corresponds to the cumulative out-of-sample predictive record where the fit 
of $y_i$ is measured with respect to the posterior density using data $\{ y_j 
\}_{j<i}$ up to the $i$th data point. This is in sharp contrast to in-sample 
measures of fit that condition on the entire data set $y$. Also, the marginal 
likelihood is invariant to permutations in the indices of the data, so that 
the same $m(y|\mathcal{M}_{s})$ will be obtained if the data are rearranged.

We next consider the computation of the marginal likelihood for the FREQ and REQ models.

%------------------------------------------------------------------------------
\subsubsection{Marginal Likelihood for the FREQ Model}\label{sec:mlFREQ}
%------------------------------------------------------------------------------
The marginal likelihood for the FREQ model is derived following 
\citet{Chib-Jeliazkov-2001} since the conditional posterior for ($\sigma, 
\gamma$) does not have a tractable form and is sampled using an MH algorithm 
(see Algorithm~\ref{alg:algorithm1}). Let $\Theta = (\beta, \Omega, 
\Theta_{1})$ where $\Theta_{1} = (\sigma,\gamma)$, then the joint posterior 
density for the FREQ model (marginally of $\alpha$, $\nu$, and $h$) can be 
expressed as,
%-----------------------------
\begin{equation}
\pi(\beta^{\ast},\Omega^{\ast},\Theta_{1}^{\ast}|y) = 
\pi(\Theta_{1}^{\ast}|y) \, \pi(\beta^{\ast}|y,\Theta_{1}^{\ast}) \, 
\pi(\Omega^{\ast}|y,\beta^{\ast},\Theta_{1}^{\ast}) ,
\label{eq:POFREQ}
\end{equation}
%-----------------------------
where $(\beta^{\ast},\Omega^{\ast}, \Theta_{1}^{\ast})$ denotes a high 
density point of $(\beta, \Omega, \Theta_{1})$. The latent variables 
$(\alpha, \nu, h)$ are marginalized to reduce the computational burden since 
computing high dimensional ordinates is costly and leads to inefficient 
estimates. Moreover, in the decomposition presented in 
Equation~\eqref{eq:POFREQ}, we have intentionally placed the intractable 
posterior ordinate $\pi(\Theta_{1}^{\ast}|y)$ first so as to avoid the MH 
step in the \emph{reduced MCMC run} -- the process of running an MCMC sampler 
with one or more parameters fixed at some value \citep{Greenberg-2012}. We 
first estimate  $\pi(\Theta_{1}^{\ast}|y)$, followed by 
$\pi(\beta^{\ast}|y,\Theta_{1}^{\ast})$, and lastly, 
$\pi(\Omega^{\ast}|y,\beta^{\ast},\Theta_{1}^{\ast})$.

To get an estimate of $\pi(\Theta_{1}^{\ast}|y)$, we first need to express the ordinate in a computationally convenient formulation. We know $\Theta_{1}$ is sampled using an MH step, which requires a proposal density and a transition kernel. Define the transition kernel from $\Theta_{1}$ to $\Theta_{1}^{\ast}$ as,
%-----------------------------
\begin{equation}
P(\Theta_{1}, \Theta_{1}^{\ast}|y,\beta,\Omega,\alpha) = 
\alpha_{MH}(\Theta_{1},\Theta_{1}^{\ast}|y,\beta,\Omega,\alpha) \; 
q(\Theta_{1}, \Theta_{1}^{\ast}|y,\beta,\Omega,\alpha),
\label{eq:transKernel}
\end{equation}
%-----------------------------
where $q(\Theta_{1}, \Theta_{1}^{\ast}|y,\beta,\Omega,\alpha)$ denotes the 
proposal density for the transition from $\Theta_{1}$ to $\Theta_{1}^{\ast}$, 
and 
%-----------------------------
\begin{equation}
\alpha_{MH}(\Theta_{1},\Theta_{1}^{\ast}|y,\beta,\Omega,\alpha) = \min 
\bigg\{ 1, \frac{f_{GAL}(y,\alpha| \beta,\Theta_{1}^{\ast} ) \; \pi(\beta, 
\Theta_{1}^{\ast})} {f_{GAL}(y,\alpha| \beta,\Theta_{1} ) \; \pi(\beta, 
\Theta_{1})}  \;
\frac{q(\Theta_{1}^{\ast}, \Theta_{1}|y,\beta,\Omega,\alpha)}{q(\Theta_{1}, 
\Theta_{1}^{\ast}|y,\beta,\Omega,\alpha)}
\bigg\},
\label{eq:alphaMH}
\end{equation}
%-----------------------------
denotes the probability of making the move from $\Theta_{1}$ to 
$\Theta_{1}^{\ast}$. Note that the conditioning of the proposal density on 
$y$ and the remaining parameters is only for the sake of generality, and a 
particular proposal density may be independent of both $y$ and $(\beta, 
\Omega, \alpha)$. Since the transition kernel, (i.e., 
Equation~\ref{eq:transKernel}) satisfies the reversibility condition, we 
exploit this property and, through suitable modifications following 
\citet{Chib-Jeliazkov-2001}, arrive at the following expression, 
%-----------------------------
\begin{equation}
\pi(\Theta_{1}^{\ast}|y) = \frac{E_{1} \{ \alpha_{MH}(\Theta_{1},
\Theta_{1}^{\ast}|y,\beta,\Omega,\alpha) \, q(\Theta_{1},
\Theta_{1}^{\ast}) | y,\beta,\Omega,\alpha) \} }{E_{2}\{ 
\alpha_{MH}(\Theta_{1}^{\ast},
\Theta_{1}|y,\beta,\Omega,\alpha)\} },
\label{eq:Theta1PostOrd}
\end{equation}
%-----------------------------
where $E_{1}$ represents expectation with respect to the posterior 
distribution $\pi(\Theta_{1},\beta,\Omega,\alpha|y)$ and $E_{2}$ 
represents expectation with respect to the distribution 
$\pi(\beta,\Omega,\alpha|y,\Theta_{1}^{\ast}) \times 
q(\Theta_{1}^{\ast},\Theta_{1}|y)$. In this formulation, the numerator in 
Equation~\eqref{eq:Theta1PostOrd} can be estimated by using draws $\{ 
\Theta_{1}^{(m)},\beta^{(m)},\Omega^{(m)},\alpha^{(m)} \}_{m=1}^{M}$ from 
the \emph{complete MCMC run} and taking an average of $\alpha_{MH}(\Theta_{1},
\Theta_{1}^{\ast}|y,\beta,\Omega,\alpha) \, q(\Theta_{1},
\Theta_{1}^{\ast}) | y,\beta,\Omega,\alpha)$, where $\alpha_{MH}(\Theta_{1},
\Theta_{1}^{\ast}|y,\beta,\Omega,\alpha)$ is given by 
Equation~\eqref{eq:alphaMH} and $q(\Theta_{1}, 
\Theta_{1}^{\ast})|y,\beta,\Omega,\alpha)$ is bivariate truncated normal 
distribution described in Algorithm~\ref{alg:algorithm1}.

To compute the denominator in Equation~\eqref{eq:Theta1PostOrd}, we note that 
the distribution $\pi(\beta,\Omega,\alpha|y,\Theta_{1}^{\ast})$ is 
conditioned on $\Theta_{1}^{\ast}$. Therefore, we conduct a \emph{reduced 
run} of Algorithm~\ref{alg:algorithm1}, i.e., sample $\beta$, $\alpha$, 
$\Omega$, $v$, and $h$ with $\Theta_{1} = (\sigma, \gamma)$ fixed at 
$\Theta_{1}^{\ast}=(\sigma^{\ast}, \gamma^{\ast})$. Additionally, at each 
iteration of the reduced run, we generate, $\Theta_{1}^{(m)} \sim 
q(\Theta_{1}^{\ast}, \Theta_{1}|y,\beta^{(m)},\Omega^{(m)}, \alpha^{(m)})$. 
The draws $\{ \beta^{(m)}, \Omega^{(m)}, \alpha^{(m)}, \Theta_{1}^{(m)} \}$ 
obtained from such a procedure are draws from 
$\pi(\beta,\Omega,\alpha|y,\Theta_{1}^{\ast}) \times 
q(\Theta_{1}^{\ast},\Theta_{1}|y)$ which can be utilized to compute the 
denominator. Therefore, an estimate of the posterior ordinate, 
$\pi(\Theta_{1}^{\ast}|y)$, can be obtained as,
%-----------------------------
\begin{equation}
\hat{\pi}(\Theta_{1}^{\ast}|y) = \frac{M^{-1} \sum_{m=1}^{M} \{ 
\alpha_{MH}(\Theta_{1}^{(m)}, \Theta_{1}^{\ast}|y,\beta^{(m)}, \Omega^{(m)}, 
\alpha^{(m)})  \; q(\Theta_{1}^{(m)}, \Theta_{1}^{\ast}|y,\beta^{(m)}, 
\Omega^{(m)}, \alpha^{(m)})\} } 
{M_{1}^{-1} \sum_{m=1}^{M_{1}}  \alpha_{MH}(\Theta_{1}^{\ast}, 
\Theta_{1}^{(m)}|y,\beta^{(m)}, \Omega^{(m)}, \alpha^{(m)}) },
\label{eq:EstTheta1PostOrd}
\end{equation}
%-----------------------------
where $M$ and $M_{1}$ denote the number of MCMC draws from the \emph{complete} and (first) \emph{reduced} MCMC runs.

Next, we need to estimate $\pi(\beta^{\ast}|y,\Theta_{1}^{\ast})$ and 
$\pi(\Omega^{\ast}|y,\beta^{\ast},\Theta_{1}^{\ast})$. We already have the 
sequence of draws $\{ \beta^{(m)}, \Omega^{(m)}, \alpha^{(m)}, \nu^{(m)}, 
h^{(m)} \}_{m=1}^{M_{1}}$ from the \emph{reduced} MCMC run conditioned on 
$\Theta_{1}^{\ast}$. These draws are utilized to estimate 
$\pi(\beta^{\ast}|y,\Theta_{1}^{\ast}) = \int \pi(\beta^{\ast} |y, 
\Theta_{1}^{\ast}, \alpha, \Omega, \nu, h) \, \pi( \alpha, \Omega, \nu, 
h|y,\Theta_{1}^{\ast}) \, d\alpha  \, d\Omega 
\, d\nu \, dh, $ as follows,
%-----------------------------
\begin{equation}
\hat{\pi}(\beta^{\ast}|y, \Theta_{1}^{\ast}) = \frac{\sum_{m=1}^{M_{1}} 
\pi(\beta^{\ast}|y, \Theta_{1}^{\ast}, \Omega^{(m)}, \alpha^{(m)}, 
\nu^{(m)}, h^{(m)}) }{M_{1}}.
\label{eq:EstbetaPostOrd}
\end{equation}
%-----------------------------

To estimate $\pi(\Omega^{\ast}|y,\beta^{\ast},\Theta_{1}^{\ast})$, we 
conduct a second \emph{reduced MCMC run}, i.e., run 
Algorithm~\ref{alg:algorithm1} for $M_{2}$ iterations with $(\beta, 
\Theta_{1})$ fixed at $(\beta^{\ast}, \Theta_{1}^{\ast})$. The resulting 
draws $\{ \alpha^{(m)}, \Omega^{(m)},  \nu^{(m)}, h^{(m)} 
\}_{m=1}^{M_{2}}$ are utilized to estimate $\pi(\Omega^{\ast}|y, 
\beta^{\ast}, \Theta_{1}^{\ast}) = \int \pi(\Omega^{\ast}| y, 
\beta^{\ast}, \Theta_{1}^{\ast}, \alpha, \nu, h) \, \pi( \alpha, \nu, 
h|y,\beta^{\ast}, \Theta_{1}^{\ast})\, d\alpha \, d\nu \, dh,$ 
as given by,
%-----------------------------
\begin{equation}
\hat{\pi}(\Omega^{\ast}|y,\beta^{\ast}, \Theta_{1}^{\ast}) = 
\frac{\sum_{m=1}^{M_{2}} 
\pi(\Omega^{\ast}|y, \beta^{\ast},  \Theta_{1}^{\ast}, \alpha^{(m)},  
\nu^{(m)}, h^{(m)}) }{M_{2}}.
\label{eq:Estvarphi2PostOrd}
\end{equation}
%-----------------------------
Substituting the expression from 
Equations~\eqref{eq:EstTheta1PostOrd}-\eqref{eq:Estvarphi2PostOrd} in 
Equation~\eqref{eq:POFREQ}, we have an estimate of the joint posterior 
ordinate $\pi(\beta^{\ast}, \Omega^{\ast}, \Theta_{1}^{\ast}|y)$.

The other quantities in the marginal likelihood (see 
Equation~\ref{eq:BMLI}) are prior ordinates and the likelihood of the FREQ 
model. Both quantities require straightforward evaluations. All the prior 
distributions are completely known (see Equation~\ref{eq:priors}), so prior 
ordinates can be easily evaluated at a chosen high-density point 
$\Theta^{\ast} = (\beta^{\ast}, \Omega^{\ast}, \sigma^{\ast}, 
\gamma^{\ast})$. The likelihood also requires evaluation at $\Theta^{\ast}$, 
but we first need to express it marginally of $(\alpha, \nu, h)$ since we 
marginalized them while computing the joint posterior ordinate. The required 
FREQ model likelihood can be written as,
%------------------------------
\begin{eqnarray*}
f(y|\beta,\Omega,\sigma,\gamma) &=& \int f(y,\alpha|\beta,\Omega, 
\sigma, \gamma ) \, d\alpha \\
& = & \int \prod_{i=1}^{n} \bigg[
\bigg\{ \prod_{j=1}^{T_{i}} f_{GAL} \left( y_{it} | x'_{it} \beta + z'_{it} \alpha_{i},
\sigma, p_{0}, \gamma \right) \bigg\} f(\alpha_{i}|\Omega) \, 
d\alpha_{i}\bigg],
\label{eq:GALfulllikelihood}
\end{eqnarray*}
%------------------------------
where $f_{GAL}$ denotes the density of GAL distribution. The likelihood can 
be computed at $\Theta^{\ast} = (\beta^{\ast}, \Omega^{\ast}, \sigma^{\ast}, 
\gamma^{\ast})$ using Monte Carlo integration as follows,
%------------------------------
\begin{equation*}
f(y|\beta^{\ast},\Omega^{\ast},\sigma^{\ast},\gamma^{\ast}) \simeq 
\sum_{j=1}^{J} \frac{ 
f(y|\beta^{\ast},\sigma^{\ast},\gamma^{\ast},\alpha^{(j)})}{J},
\end{equation*}
%------------------------------
where $\{\alpha_{i}^{(j)}\}$ are draws from $f(\alpha_{i}|\Omega^{\ast})$ 
for $i=1,\cdots,n$, and $J$ is some large number. Additionally, $(\nu,h)$ are 
automatically marginalized since $f_{GAL}(\cdot)$ is the GAL density that 
does not involve any mixture variables \citep[see Equation~(2) in][for the 
form of the density]{Rahman-Karnawat-2019}.

%------------------------------------------------------------------------------
\subsubsection{Marginal Likelihood for the REQ Model}\label{sec:mlREQ}
%------------------------------------------------------------------------------

The derivation of the marginal likelihood for REQ model follows 
\citet{Chib-1995} since all the conditional posteriors have a known form (see 
Algorithm~\ref{alg:algorithm2}). Let $\Theta = (\beta, \sigma, \Omega)$, then 
the joint posterior (marginally of $\alpha$ and $\nu$) can be expressed as,
%--------------------------
\begin{equation}
    \pi(\Theta^{\ast}|y) = \pi(\beta^{\ast}|y) \, 
    \pi(\Omega^{\ast}|y,\beta^{\ast}) \, \pi(\sigma^{\ast}|y,\beta^{\ast}, 
    \Omega^{\ast}), 
    \label{eq:POREQ}
\end{equation}
%-------------------------- 
where the $\ast$ on the parameters denotes a high-density point. Each expression on the right-hand side of Equation~\eqref{eq:POREQ} can be written in terms of the conditional posteriors (see Algorithm~\ref{alg:algorithm2}) and an estimate is obtained by taking the ergodic average of the conditional posterior density with MCMC draws either from the complete or reduced runs. 

The posterior density $\pi(\beta^{\ast}|y)$ is expressed as 
$\pi(\beta^{\ast}|y) = \int \pi(\beta^{\ast}|y,\nu, \sigma, \Omega) \, 
\pi(\nu,\sigma,\Omega|y) \, d\nu 
\, d\sigma \, d\Omega$ and its estimate is computed as $ 
\hat{\pi}(\beta^{\ast}|y) = G^{-1} \sum_{g=1}^{G} 
\pi(\beta^{\ast}|y,\nu^{(g)},\sigma^{(g)}, \Omega^{(g)})$, where the G 
draws are from the \emph{complete} MCMC run. The remaining two terms are 
reduced conditional density ordinates and require MCMC draws from two 
separate \emph{reduced runs}. To obtain an estimate of 
$\pi(\Omega^{\ast}|y,\beta^{\ast}) = \int 
\pi(\Omega^{\ast}|y,\beta^{\ast},\sigma,\alpha, \nu) \, 
\pi(\sigma,\alpha,\nu|y,\beta^{\ast}) \, d\sigma \, d\alpha 
\, d\nu$, we conduct a (first) \emph{reduced run}, i.e., run 
Algorithm~\ref{alg:algorithm2} for $G_{1}$ iterations with $\beta$ fixed at 
$\beta^{\ast}$. We then compute an estimate of the ordinate as $ 
\hat{\pi}(\Omega^{\ast}|y,\beta^{\ast}) = G_{1}^{-1} \sum_{g=1}^{G_{1}} 
\pi(\Omega^{\ast}|y,\beta^{\ast},\sigma^{(g)},\alpha^{(g)},\nu^{(g)})$.
Finally, an estimate of the third term $\pi(\sigma^{\ast}|y, \beta^{\ast}, 
\Omega^{\ast}) = \int \pi(\sigma^{\ast}|y, \beta^{\ast}, \Omega^{\ast}, 
\alpha, \nu) \, \pi(\alpha,\nu|y,\beta^{\ast},\Omega^{\ast}) \, d\alpha \, 
d\nu$, is obtained as $ 
\hat{\pi}(\sigma^{\ast}|y,\beta^{\ast}, \Omega^{\ast}) = G_{2}^{-1} 
\sum_{g=1}^{G_{2}} \pi(\sigma^{\ast}|y,\beta^{\ast},\Omega^{\ast}, 
\alpha^{(g)},\nu^{(g)})$, where the $G_{2}$ Gibbs draws are from the second 
\emph{reduced run} of Algorithm~\ref{alg:algorithm2} with $(\beta, \Omega)$ 
fixed at $(\beta^{\ast}, \Omega^{\ast})$.

With an estimate of the joint posterior ordinate now available, we need to 
compute the prior ordinates and the likelihood to estimate the marginal 
likelihood for REQ model. The prior ordinates are readily available since the 
prior distributions for ($\beta, \sigma, \Omega$) have tractable forms. 
The likelihood is calculated marginally of ($\alpha, \nu$) since we 
marginalized them while computing the joint posterior ordinate. The required 
likelihood can be written as,
%------------------------------
\begin{equation*}
f(y|\beta,\sigma,\Omega) = \int f(y,\alpha|\beta,\sigma, \Omega) \, d\alpha = 
\int \prod_{i=1}^{n} \bigg[
\bigg\{ \prod_{t=1}^{T_{i}} f_{AL} \left( y_{it} | x'_{it} \beta + z'_{it} 
\alpha_{i},
\sigma, p \right) \bigg\} f(\alpha_{i}|\Omega) \, d\alpha_{i}\bigg],
\label{eq:ALfulllikelihood}
\end{equation*}
%------------------------------
where $f_{AL}$ denotes the density of the AL distribution. The above 
expression can be computed using Monte Carlo integration at 
$\Theta^{\ast}=(\beta^{\ast},\sigma^{\ast},\Omega^{\ast})$ as follows, 
%------------------------------
\begin{equation*}
f(y|\beta^{\ast},\sigma^{\ast},\Omega^{\ast}) \simeq \sum_{j=1}^{J} \frac{ 
f(y|\beta^{\ast},\sigma^{\ast}, \alpha^{(j)})}{J},
\end{equation*}
%------------------------------
where $\{\alpha_{i}^{(j)}\}$ are draws from $f(\alpha_{i}|\Omega^{\ast})$ 
for $i=1,\cdots,n$, with $J$ being a large number. Note that $\nu$ is 
automatically marginalized by virtue of not using the mixture representation 
of AL distribution in the likelihood. 

%------------------------------------------------------------------------------
\section{Simulation Studies}\label{sec:simStudies}
%------------------------------------------------------------------------------

In this section, we conduct multiple simulation studies to illustrate the 
performance of the proposed MCMC algorithm for estimating the FREQ model, 
estimated using Algorithm~\ref{alg:algorithm1}, and 
compare the results to those from the REQ model, estimated using 
Algorithm~\ref{alg:algorithm2}. Moreover, we compute the marginal likelihood 
for both the models to examine the benefits (i.e., better model fitting), if 
any, of the FREQ model \emph{vis-a-vis} the REQ model. The data for the 
simulation studies are generated from the following panel data model,
%---------------------------------------------
\begin{equation*}
y_{it} = \alpha_{1i} + \alpha_{2i} \, z_{2it} + \beta_{1} + \beta_{2} \, 
x_{2it} + \beta_{3} \, x_{3it} + \varepsilon_{it},
\end{equation*}
%---------------------------------------------
where $\alpha = (\alpha_{1}, \alpha_{2})' \sim N_{2}( [0,0]', [1, 0; 0, 
1])$, $\beta = (\beta_{1}, \beta_{2}, \beta_{3})' = (10, 5, 2 )'$, $z_{2it} 
\sim \mathrm{Unif}(1,3)$, $x_{2} \sim N(0,0.25)$, $x_{3} \sim N(2,0.25)$, 
and the errors $\varepsilon$ were generated from a standard logistic 
distribution $\mathcal{L}(0,1)$. We generate 9 different data samples with 
$T_{i}= (5, 10, 15)$ and $n=(100,250,500)$, where $T_{i}$ denotes the number 
of repeated observations for each individual $i$ and $n$ represents the 
number of individuals.

\begin{figure*}[!b]
	\centerline{
		\mbox{\includegraphics[width=7.0in, height=6.0in]{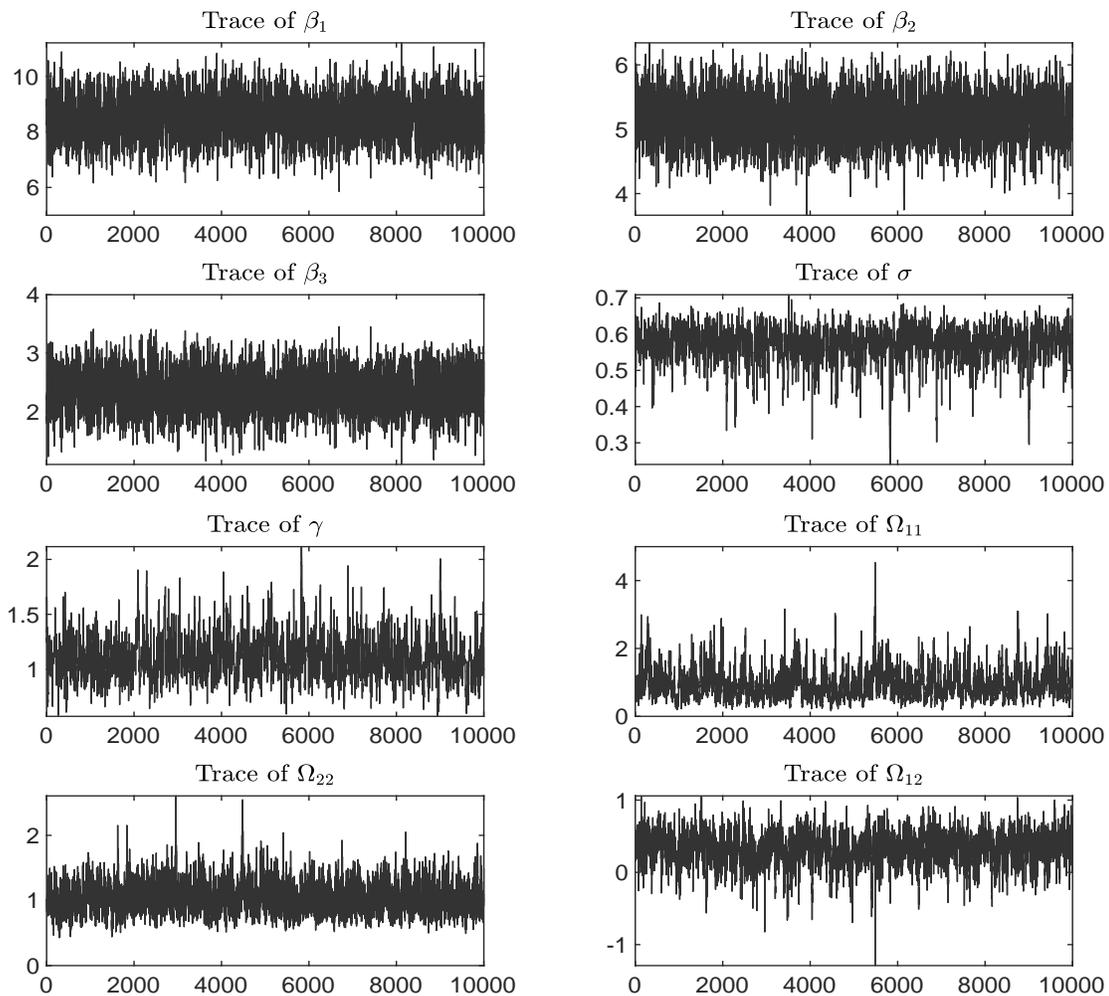}} }
	\caption{Trace plots of the MCMC draws for Simulation Study~1 ($m=5, 
		n=100$)
		at the 25th quantile.} \label{fig:traceplots}
\end{figure*}

In each simulation study, the posterior estimates of the parameters in the
FREQ model are obtained based on the simulated data and the following prior
distributions: $\beta \sim N(0_{k}, 100 I_{k})$, $\Omega
\sim IW(\omega_{0}, O_{0})$, $\sigma \sim 
IG(5/2,8/2)$, and $\gamma \sim
\mathrm{Unif}(L,U)$. Here, $\omega_{0}=5+l$, $O_{0}= (\omega_{0} - l-1)\ast 
I_{l}$, and $(L,U)$ are obtained as mentioned in
Section~\ref{sec:meth}. The same prior distributions are employed for the REQ model. Table~\ref{Table:SimResult} reports, for each simulated dataset, the MCMC results at five different quantiles obtained from 10,000 iterations after a burn-in of 2,500 iterations. Inefficiency factors are calculated using the formula,
%---------------------------------
$
1 + 2 \sum_{t=1}^{T} \rho_{k}(t) \bigg( \frac{T-t}{T}  \bigg),
$
%---------------------------------
where $\rho_{k}(t)$ denotes the autocorrelation for the $k$th parameter at
lag $t$, and $T$ is the value at which the autocorrelations taper off 
(typically, 0.05 or 0.10). In the MH sampling of $(\sigma, \gamma)$, the 
tuning factor $\iota$ is adjusted to obtain an acceptance rate of 
approximately 30 percent. Convergence of the MCMC draws is quick, as 
demonstrated in the trace plots for the 25th quantile from Simulation Study~1 
in Figure~\ref{fig:traceplots}. The trace plots for the remaining quantiles 
in Simulation Study~1 and all quantiles in the other 8 simulation studies are 
similar and reflect quick convergence to the joint posterior distribution.

%---------------------------  Figure 2 ---------------------------------------
%------------------------------------------------------------------------------

%------------------------------- Table 1 -------------------------------------
\begin{footnotesize}
{\setlength{\tabcolsep}{3.5pt}
\setlength{\extrarowheight}{1.2pt}
\setlength\LTcapwidth{\linewidth}
\begin{longtable}{c rrr rrr rrr rrr rrr rrr rr}
\caption{Posterior mean (\textsc{mean}), standard deviation (\textsc{sd})
and inefficiency factor (\textsc{if}) of the parameters in the family of FREQ models from nine simulation studies: SS1 ($T_{i}=5,n=100$), SS2 ($T_{i}=5,n=250$), SS3 ($T_{i}=5,n=500$), SS4 ($T_{i}=10,n=250$), SS5 ($T_{i}=10,n=250$), SS6 ($T_{i}=10,n=500$), SS7 ($T_{i}=15,n=100$), SS8 ($T_{i}=15,n=250$, and SS9 ($T_{i}=15,n=500$))} \\
%\begin{tabular}
\toprule
& & \multicolumn{3}{c}{\textsc{10th qtl}}
& & \multicolumn{3}{c}{\textsc{25th qtl}}
& & \multicolumn{3}{c}{\textsc{50th qtl}}
& & \multicolumn{3}{c}{\textsc{75th qtl}} 
& & \multicolumn{3}{c}{\textsc{90th qtl}}\\
\cmidrule{3-5} \cmidrule{7-9}  \cmidrule{11-13} \cmidrule{15-17}  \cmidrule{19-21}
%------------------------------------------------------------------------------
\endfirsthead
\multicolumn{21}{c}%
{\tablename\ \thetable\ -- \textit{Continued from previous page}} \\
\hline
\endhead
\endfoot
%------------------------------------------------------------------------------
%                   (m=5, n=100, 250, 500)
%------------------------------------------------------------------------------
\textsc{SS1} & &  \textsc{mean} & \textsc{sd} & \textsc{if}
& &  \textsc{mean} & \textsc{sd} & \textsc{if}
& &  \textsc{mean} & \textsc{sd} & \textsc{if}
& &  \textsc{mean} & \textsc{sd} & \textsc{if}
& &  \textsc{mean} & \textsc{sd} & \textsc{if}   \\
\midrule
%------------------------------------------------------------------------------
$\beta_{1}$   &&  7.39  & 0.72  &  4.47  &&   8.51  & 0.71  &  3.03   
              &&  9.41  & 0.74  &  2.46  &&  10.58  & 0.74  &  3.69    
              && 11.60  & 0.75  &  5.15\\
$\beta_{2}$   &&  5.11  & 0.35  &  5.21  &&   5.17  & 0.34  &  3.43   
              &&  5.25  & 0.33  &  2.41  &&   5.15  & 0.35  &  4.00    
              &&  4.95  & 0.38  &  5.86\\
%-----------------------------------------------------------------------------
$\beta_{3}$   &&  2.42  & 0.33  &  4.54  &&   2.34  & 0.33  &  3.11   
              &&  2.36  & 0.34  &  2.33  &&   2.31  & 0.35  &  3.82    
              &&  2.29  & 0.35  &  5.31\\
$\sigma$      &&  0.44  & 0.03  &  8.24  &&   0.57  & 0.05  & 10.02   
              &&  0.65  & 0.03  &  8.05  &&   0.53  & 0.06  &  9.43    
              &&  0.44  & 0.03  &  7.95\\
%-----------------------------------------------------------------------------
$\gamma$      &&  2.81  & 0.32  &  8.81  &&   1.11  & 0.21  & 11.59   
	     	  &&$-0.10$ & 0.09  & 13.22  && $-1.28$ & 0.20  &  8.97    
			  &&$-2.90$ & 0.28  &  6.56\\
$\Omega_{11}$ &&  1.09  & 0.51  & 16.62  &&   0.92  & 0.41  & 16.53   
 			  &&  1.08  & 0.52  & 15.30  &&   0.88  & 0.39  & 15.24    
			  &&  0.73  & 0.35  & 15.12\\
%-----------------------------------------------------------------------------
$\Omega_{22}$ &&  0.97  & 0.22  & 10.25  &&   1.02  & 0.22  & 10.91   
              &&  0.99  & 0.22  & 10.23  &&   1.04  & 0.23  & 13.25    
              &&  1.14  & 0.24  & 13.50\\
$\Omega_{12}$ &&  0.36  & 0.22  &  7.98  &&   0.31  & 0.21  &  7.78   
              &&  0.32  & 0.22  &  6.57  &&   0.28  & 0.23  &  8.41    
              &&  0.17  & 0.23  &  7.96\\
%-----------------------------------------------------------------------------
\midrule \textsc{SS2} 
& & \multicolumn{3}{c}{\textsc{10th qtl}} & & \multicolumn{3}{c}{\textsc{25th qtl}} 
& & \multicolumn{3}{c}{\textsc{50th qtl}} & & \multicolumn{3}{c}{\textsc{75th qtl}} 
& & \multicolumn{3}{c}{\textsc{90th qtl}}\\
\midrule
%------------------------------------------------------------------------------
$\beta_{1}$   &&  7.21  & 0.48  &  5.11  &&   8.11  & 0.48  &  4.15   
			  &&  9.19  & 0.45  &  2.14  &&  10.29  & 0.47  &  4.64    
			  && 11.23  & 0.48  &  6.04\\
$\beta_{2}$   &&  4.87  & 0.24  &  5.67  &&   4.89  & 0.24  &  4.56   
			  &&  4.92  & 0.23  &  2.22  &&   4.89  & 0.24  &  5.05    
			  &&  4.85  & 0.23  &  5.64\\
%-----------------------------------------------------------------------------
$\beta_{3}$   &&  2.46  & 0.23  &  5.34  &&   2.50  & 0.23  &  4.24   
			  &&  2.48  & 0.21  &  2.17  &&   2.45  & 0.22  &  4.73    
			  &&  2.47  & 0.23  &  6.22\\
$\sigma$      &&  0.43  & 0.03  & 10.80  &&   0.51  & 0.07  & 18.22   
			  &&  0.67  & 0.02  &  7.14  &&   0.48  & 0.07  & 23.21    
			  &&  0.43  & 0.02  &  9.66\\
%-----------------------------------------------------------------------------
$\gamma$      &&  3.02  & 0.21  &  9.36  &&   1.41  & 0.20  & 16.94   
			  &&  0.01  & 0.06  & 16.17  && $-1.50$ & 0.21  & 21.92    
			  &&$-3.02$ & 0.21  &  8.55\\
$\Omega_{11}$ &&  1.76  & 0.78  & 54.04  &&   1.00  & 0.49  & 37.87   
			  &&  1.39  & 0.66  & 42.82  &&   0.95  & 0.43  & 58.22    
			  &&  1.26  & 0.59  & 50.67\\
%-----------------------------------------------------------------------------
$\Omega_{22}$ &&  1.18  & 0.19  & 49.19  &&   1.03  & 0.16  & 33.04   
			  &&  1.11  & 0.19  & 38.52  &&   1.05  & 0.16  & 37.59    
			  &&  1.04  & 0.16  & 32.66\\
$\Omega_{12}$ &&$-0.31$ & 0.32  & 17.18  &&   0.04  & 0.24  & 16.56   
			  &&$-0.13$ & 0.31  & 17.28  &&   0.03  & 0.21  & 10.31    
			  &&  0.02  & 0.25  & 13.03\\
%-----------------------------------------------------------------------------
\midrule \textsc{SS3} 
& & \multicolumn{3}{c}{\textsc{10th qtl}} & & \multicolumn{3}{c}{\textsc{25th qtl}} 
& & \multicolumn{3}{c}{\textsc{50th qtl}} & & \multicolumn{3}{c}{\textsc{75th qtl}} 
& & \multicolumn{3}{c}{\textsc{90th qtl}}\\
\midrule
%-----------------------------------------------------------------------------
$\beta_{1}$   &&  8.10  & 0.32  &  5.23  &&   9.00  & 0.32  &  3.69   
			  &&  9.98  & 0.32  &  2.25  &&  11.06  & 0.33  &  3.71    
			  && 11.93  & 0.33  &  6.26\\
$\beta_{2}$   &&  4.85  & 0.16  &  5.42  &&   4.78  & 0.16  &  3.70   
			  &&  4.78  & 0.16  &  2.32  &&   4.78  & 0.16  &  3.79    
			  &&  4.75  & 0.16  &  5.62\\
%-----------------------------------------------------------------------------
$\beta_{3}$   &&  1.94  & 0.15  &  5.40  &&   1.97  & 0.15  &  3.81   
			  &&  1.97  & 0.15  &  2.47  &&   1.97  & 0.16  &  3.81    
			  &&  2.04  & 0.16  &  6.43\\
$\sigma$      &&  0.44  & 0.02  &  9.47  &&   0.55  & 0.04  & 16.73   
			  &&  0.67  & 0.02  &  6.52  &&   0.55  & 0.04  & 12.82    
			  &&  0.44  & 0.02  &  8.45\\
%-----------------------------------------------------------------------------
$\gamma$      &&  2.94  & 0.15  &  9.24  &&   1.27  & 0.13  & 16.42   
			  &&$-0.04$ & 0.04  & 16.34  && $-1.27$ & 0.12  & 12.65    
			  &&$-2.95$ & 0.15  &  8.53\\
$\Omega_{11}$ &&  0.92  & 0.34  & 62.18  &&   0.73  & 0.26  & 44.29   
			  &&  0.95  & 0.33  & 53.71  &&   0.69  & 0.26  & 49.39    
			  &&  0.82  & 0.35  & 54.32\\
%-----------------------------------------------------------------------------
$\Omega_{22}$ &&  1.05  & 0.12  & 37.81  &&   1.06  & 0.12  & 25.28   
			  &&  1.06  & 0.12  & 31.33  &&   1.05  & 0.11  & 26.68    
			  &&  1.13  & 0.13  & 39.60\\
$\Omega_{12}$ &&  0.19  & 0.16  & 16.11  &&   0.19  & 0.15  & 13.56   
			  &&  0.15  & 0.16  & 10.41  &&   0.20  & 0.14  & 12.79    
			  &&  0.06  & 0.17  & 17.26\\
%------------------------------------------------------------------------------
%                   (m=10, n=100, 250, 500)
%------------------------------------------------------------------------------
\midrule \textsc{SS4} 
& & \multicolumn{3}{c}{\textsc{10th qtl}} & & \multicolumn{3}{c}{\textsc{25th qtl}} 
& & \multicolumn{3}{c}{\textsc{50th qtl}} & & \multicolumn{3}{c}{\textsc{75th qtl}} 
& & \multicolumn{3}{c}{\textsc{90th qtl}}\\
\midrule
%-----------------------------------------------------------------------------
$\beta_{1}$   &&  8.04  & 0.49  &  5.99  &&   8.89  & 0.49  &  4.80   
			  &&  9.84  & 0.49  &  2.64  &&  10.84  & 0.49  &  4.79    
			  && 11.67  & 0.50  &  5.98\\
$\beta_{2}$   &&  4.85  & 0.23  &  6.83  &&   4.86  & 0.23  &  5.33   
			  &&  4.84  & 0.23  &  2.99  &&   4.79  & 0.22  &  4.37    
			  &&  4.69  & 0.24  &  6.46\\
%-----------------------------------------------------------------------------
$\beta_{3}$   &&  2.09  & 0.23  &  6.52  &&   2.09  & 0.23  &  5.08   
			  &&  2.13  & 0.23  &  2.79  &&   2.16  & 0.23  &  5.01    
			  &&  2.19  & 0.23  &  6.53\\
$\sigma$      &&  0.44  & 0.02  &  7.89  &&   0.52  & 0.07  & 11.03   
			  &&  0.68  & 0.02  &  6.28  &&   0.53  & 0.06  & 12.68    
			  &&  0.44  & 0.02  &  8.34\\
%-----------------------------------------------------------------------------
$\gamma$      &&  2.98  & 0.21  &  8.91  &&   1.38  & 0.20  & 10.45   
			  &&  0.02  & 0.06  & 13.06  && $-1.34$ & 0.18  & 12.00    
			  &&$-2.92$ & 0.20  &  7.44\\
$\Omega_{11}$ &&  0.74  & 0.33  & 16.67  &&   0.77  & 0.35  & 20.20   
			  &&  0.97  & 0.47  & 16.50  &&   0.86  & 0.41  & 22.23    
			  &&  1.06  & 0.50  & 29.58\\
%-----------------------------------------------------------------------------
$\Omega_{22}$ &&  0.81  & 0.17  & 12.97  &&   0.83  & 0.17  & 11.74   
			  &&  0.86  & 0.17  & 11.46  &&   0.84  & 0.17  & 15.68    
			  &&  0.92  & 0.18  & 15.91\\
$\Omega_{12}$ &&  0.02  & 0.18  &  6.59  && $-0.01$ & 0.19  &  5.70   
		      &&$-0.10$ & 0.21  &  4.39  && $-0.06$ & 0.21  &  6.08    
			  &&$-0.17$ & 0.24  &  6.01\\
%-----------------------------------------------------------------------------
\midrule \textsc{SS5} 
& & \multicolumn{3}{c}{\textsc{10th qtl}} & & \multicolumn{3}{c}{\textsc{25th qtl}} 
& & \multicolumn{3}{c}{\textsc{50th qtl}} & & \multicolumn{3}{c}{\textsc{75th qtl}} 
& & \multicolumn{3}{c}{\textsc{90th qtl}}\\
\midrule
%-----------------------------------------------------------------------------
$\beta_{1}$   &&  7.97  & 0.33  &  6.27  &&   9.12  & 0.32  &  4.63   
			  && 10.29  & 0.32  &  2.59  &&  11.22  & 0.32  &  4.53    
			  && 12.20  & 0.33  &  6.48\\
$\beta_{2}$   &&  4.79  & 0.15  &  8.17  &&   4.77  & 0.14  &  4.51   
			  &&  4.75  & 0.14  &  2.67  &&   4.81  & 0.14  &  4.27    
			  &&  4.82  & 0.15  &  6.73\\
%-----------------------------------------------------------------------------
$\beta_{3}$   &&  1.99  & 0.15  &  6.56  &&   1.88  & 0.15  &  4.63   
			  &&  1.81  & 0.15  &  2.66  &&   1.86  & 0.15  &  4.72    
			  &&  1.85  & 0.15  &  6.79\\
$\sigma$      &&  0.43  & 0.01  &  7.86  &&   0.52  & 0.04  & 16.78   
			  &&  0.67  & 0.01  &  5.80  &&   0.54  & 0.03  & 10.80    
			  &&  0.44  & 0.01  &  7.59\\
%-----------------------------------------------------------------------------
$\gamma$      &&  2.94  & 0.13  &  7.02  &&   1.37  & 0.14  & 17.52   
			  &&  0.01  & 0.04  & 12.62  && $-1.30$ & 0.11  & 11.37    
			  &&$-2.92$ & 0.13  &  8.00\\
$\Omega_{11}$ &&  1.15  & 0.36  & 29.00  &&   0.91  & 0.29  & 27.29   
			  &&  1.37  & 0.34  & 11.67  &&   0.92  & 0.29  & 22.60    
			  &&  0.91  & 0.31  & 38.42\\
%-----------------------------------------------------------------------------
$\Omega_{22}$ &&  0.87  & 0.11  & 20.20  &&   0.83  & 0.11  & 16.44   
			  &&  0.89  & 0.11  &  8.46  &&   0.85  & 0.11  & 15.26    
			  &&  0.86  & 0.11  & 19.06\\
$\Omega_{12}$ &&  0.00  & 0.16  &  6.30  &&   0.09  & 0.13  &  5.53   
			  &&$-0.08$ & 0.15  &  3.42  &&   0.06  & 0.13  &  5.09    
			  &&  0.07  & 0.13  &  6.82\\
%-----------------------------------------------------------------------------
\midrule \textsc{SS6} 
& & \multicolumn{3}{c}{\textsc{10th qtl}} & & \multicolumn{3}{c}{\textsc{25th qtl}} 
& & \multicolumn{3}{c}{\textsc{50th qtl}} & & \multicolumn{3}{c}{\textsc{75th qtl}} 
& & \multicolumn{3}{c}{\textsc{90th qtl}}\\
\midrule
%-----------------------------------------------------------------------------
$\beta_{1}$   &&  8.15  & 0.23  &  6.60  &&   9.17  & 0.22  &  4.25   
			  && 10.14  & 0.21  &  2.45  &&  11.15  & 0.22  &  3.94    
			  && 12.21  & 0.22  &  6.08\\
$\beta_{2}$   &&  5.08  & 0.10  &  6.38  &&   5.06  & 0.10  &  4.48   
			  &&  5.06  & 0.10  &  2.69  &&   5.08  & 0.10  &  4.33    
			  &&  5.11  & 0.10  &  7.32\\
%-----------------------------------------------------------------------------
$\beta_{3}$   &&  1.97  & 0.10  &  7.31  &&   1.93  & 0.10  &  4.48   
			  &&  1.95  & 0.10  &  2.56  &&   1.95  & 0.10  &  4.13    
			  &&  1.89  & 0.10  &  6.47\\
$\sigma$      &&  0.42  & 0.01  &  8.24  &&   0.51  & 0.04  & 15.97   
			  &&  0.66  & 0.01  &  6.18  &&   0.53  & 0.03  & 18.51    
			  &&  0.42  & 0.01  &  8.67\\
%-----------------------------------------------------------------------------
$\gamma$      &&  2.96  & 0.09  &  7.88  &&   1.37  & 0.11  & 15.59   
			  &&  0.01  & 0.03  & 10.70  && $-1.30$ & 0.10  & 17.56    
			  &&$-2.95$ & 0.10  &  8.22\\
$\Omega_{11}$ &&  1.42  & 0.27  & 20.71  &&   1.01  & 0.24  & 37.81   
			  &&  1.26  & 0.25  & 17.14  &&   0.96  & 0.25  & 38.18    
			  &&  1.40  & 0.29  & 33.39\\
%-----------------------------------------------------------------------------
$\Omega_{22}$ &&  0.95  & 0.09  & 16.69  &&   0.91  & 0.08  & 25.55   
			  &&  0.96  & 0.09  & 12.01  &&   0.93  & 0.08  & 22.02    
			  &&  0.99  & 0.09  & 22.41\\
$\Omega_{12}$ &&$-0.10$ & 0.12  &  7.11  &&   0.03  & 0.11  &  5.58   
			  &&$-0.08$ & 0.11  &  4.07  &&   0.01  & 0.11  &  5.71    
			  &&$-0.13$ & 0.13  &  7.34\\
%------------------------------------------------------------------------------
%                   (m=15, n=100, 250, 500)
%------------------------------------------------------------------------------
\midrule \textsc{SS7} 
& & \multicolumn{3}{c}{\textsc{10th qtl}} & & \multicolumn{3}{c}{\textsc{25th qtl}} 
& & \multicolumn{3}{c}{\textsc{50th qtl}} & & \multicolumn{3}{c}{\textsc{75th qtl}} 
& & \multicolumn{3}{c}{\textsc{90th qtl}}\\
\midrule
%-----------------------------------------------------------------------------
$\beta_{1}$   &&  7.51  & 0.40  &  6.64  &&   8.59  & 0.38  &  5.55   
			  &&  9.71  & 0.40  &  2.73  &&  10.67  & 0.40  &  5.20    
			  && 11.67  & 0.41  &  6.91\\
$\beta_{2}$   &&  5.19  & 0.18  &  8.03  &&   5.14  & 0.18  &  5.96   
			  &&  5.05  & 0.18  &  2.83  &&   5.14  & 0.18  &  5.83    
			  &&  5.20  & 0.18  &  7.09\\
%-----------------------------------------------------------------------------
$\beta_{3}$   &&  2.24  & 0.18  &  7.57  &&   2.21  & 0.18  &  5.89   
			  &&  2.14  & 0.18  &  3.10  &&   2.18  & 0.18  &  6.02    
			  &&  2.19  & 0.18  &  7.45\\
$\sigma$      &&  0.42  & 0.02  &  6.37  &&   0.46  & 0.06  & 13.88   
			  &&  0.66  & 0.02  &  6.28  &&   0.49  & 0.06  & 15.42    
			  &&  0.43  & 0.02  &  7.98\\
%-----------------------------------------------------------------------------
$\gamma$      &&  3.00  & 0.16  &  6.44  &&   1.53  & 0.17  & 13.30   
			  &&$-0.01$ & 0.05  & 12.86  && $-1.42$ & 0.19  & 15.69    
			  &&$-2.98$ & 0.17  &  8.25\\
$\Omega_{11}$ &&  1.42  & 0.45  & 14.41  &&   1.01  & 0.34  & 13.40   
			  &&  1.08  & 0.36  & 10.17  &&   1.07  & 0.36  & 12.31    
			  &&  1.04  & 0.35  & 12.93\\
%-----------------------------------------------------------------------------
$\Omega_{22}$ &&  0.86  & 0.15  &  8.10  &&   0.85  & 0.15  &  6.23   
			  &&  0.87  & 0.15  &  5.68  &&   0.86  & 0.15  &  6.76    
			  &&  0.87  & 0.15  &  8.56\\
$\Omega_{12}$ &&  0.24  & 0.17  &  4.03  &&   0.30  & 0.15  &  3.33   
			  &&  0.28  & 0.16  &  2.45  &&   0.28  & 0.16  &  2.94    
			  &&  0.25  & 0.16  &  3.87\\
%-----------------------------------------------------------------------------
\midrule \textsc{SS8} 
& & \multicolumn{3}{c}{\textsc{10th qtl}} & & \multicolumn{3}{c}{\textsc{25th qtl}} 
& & \multicolumn{3}{c}{\textsc{50th qtl}} & & \multicolumn{3}{c}{\textsc{75th qtl}} 
& & \multicolumn{3}{c}{\textsc{90th qtl}}\\
\midrule
%-----------------------------------------------------------------------------
$\beta_{1}$   &&  7.79  & 0.27  &  7.72  &&   9.00  & 0.27  &  4.64   
			  && 10.13  & 0.25  &  2.65  &&  11.09  & 0.26  &  5.38    
			  && 11.99  & 0.27  &  7.86\\
$\beta_{2}$   &&  4.94  & 0.12  &  7.03  &&   4.95  & 0.12  &  4.30   
			  &&  4.95  & 0.12  &  2.82  &&   4.95  & 0.12  &  5.40    
			  &&  4.94  & 0.13  &  7.20\\
%-----------------------------------------------------------------------------
$\beta_{3}$   &&  2.02  & 0.12  &  8.45  &&   1.92  & 0.12  &  4.79   
			  &&  1.87  & 0.11  &  2.73  &&   1.92  & 0.12  &  5.67    
			  &&  1.98  & 0.12  &  8.56\\
$\sigma$      &&  0.45  & 0.01  &  7.63  &&   0.57  & 0.02  &  7.69   
			  &&  0.68  & 0.01  &  6.83  &&   0.51  & 0.04  & 12.52    
			  &&  0.44  & 0.01  &  7.43\\
%-----------------------------------------------------------------------------
$\gamma$      &&  2.90  & 0.10  &  6.25  &&   1.22  & 0.08  &  8.09   
			  &&$-0.05$ & 0.03  & 11.25  && $-1.43$ & 0.12  & 11.87    
			  &&$-2.96$ & 0.10  &  7.04\\
$\Omega_{11}$ &&  0.96  & 0.27  & 18.49  &&   0.89  & 0.22  & 21.28   
			  &&  1.13  & 0.26  & 12.34  &&   0.88  & 0.24  & 21.64    
			  &&  0.98  & 0.27  & 25.79\\
%-----------------------------------------------------------------------------
$\Omega_{22}$ &&  1.08  & 0.12  & 13.64  &&   1.02  & 0.12  & 12.89   
			  &&  1.04  & 0.12  &  8.69  &&   1.02  & 0.12  & 13.24    
			  &&  0.98  & 0.11  & 17.43\\
$\Omega_{12}$ &&  0.02  & 0.13  &  3.92  &&   0.12  & 0.12  &  3.77   
			  &&  0.06  & 0.14  &  3.09  &&   0.11  & 0.13  &  3.91    
		      &&  0.12  & 0.13  &  4.93\\
%-----------------------------------------------------------------------------
\midrule \textsc{SS9} 
& & \multicolumn{3}{c}{\textsc{10th qtl}} & & \multicolumn{3}{c}{\textsc{25th qtl}} 
& & \multicolumn{3}{c}{\textsc{50th qtl}} & & \multicolumn{3}{c}{\textsc{75th qtl}} 
& & \multicolumn{3}{c}{\textsc{90th qtl}}\\
\midrule
%-----------------------------------------------------------------------------
$\beta_{1}$   &&  8.10  & 0.19  &  6.84  &&   9.17  & 0.19  &  4.17   
			  && 10.22  & 0.18  &  2.67  &&  11.25  & 0.18  &  4.43    
			  && 12.26  & 0.19  &  6.50\\
$\beta_{2}$   &&  5.05  & 0.09  &  7.08  &&   5.06  & 0.09  &  4.16   
			  &&  5.08  & 0.08  &  3.07  &&   5.05  & 0.08  &  4.53    
			  &&  5.04  & 0.09  &  6.39\\
%-----------------------------------------------------------------------------
$\beta_{3}$   &&  1.97  & 0.09  &  7.20  &&   1.94  & 0.09  &  4.41   
			  &&  1.94  & 0.08  &  2.80  &&   1.94  & 0.08  &  4.54    
			  &&  1.92  & 0.08  &  7.01\\
$\sigma$      &&  0.46  & 0.01  &  6.88  &&   0.58  & 0.01  &  6.70   
			  &&  0.69  & 0.01  &  6.11  &&   0.56  & 0.02  &  7.00    
			  &&  0.45  & 0.01  &  7.35\\
%-----------------------------------------------------------------------------
$\gamma$      &&  2.89  & 0.07  &  7.09  &&   1.20  & 0.05  &  9.08   
			  &&$-0.00$ & 0.02  &  9.94  && $-1.28$ & 0.06  &  7.30    
			  &&$-2.89$ & 0.07  &  6.90\\
$\Omega_{11}$ &&  0.78  & 0.18  & 28.03  &&   0.78  & 0.16  & 19.30   
			  &&  1.02  & 0.16  &  9.35  &&   0.73  & 0.16  & 19.95    
			  &&  0.93  & 0.18  & 22.30\\
%-----------------------------------------------------------------------------
$\Omega_{22}$ &&  1.11  & 0.09  & 17.75  &&   1.11  & 0.09  & 11.27   
			  &&  1.16  & 0.09  &  6.93  &&   1.12  & 0.09  & 11.78    
			  &&  1.11  & 0.09  & 16.10\\
$\Omega_{12}$ &&$-0.01$ & 0.09  &  3.85  && $-0.02$ & 0.09  &  3.70   
			  &&$-0.13$ & 0.09  &  2.67  && $-0.02$ & 0.09  &  3.78    
			  &&$-0.04$ & 0.10  &  4.53\\
%-----------------------------------------------------------------------------
\bottomrule
\label{Table:SimResult}
\end{longtable}
}
\end{footnotesize}
%------------------------------------------------------------------------------

The results, presented in Table~\ref{Table:SimResult}, show that the posterior
estimates of the regression coefficients $\beta$ are close to the true values
$(10,5,2)$ with small standard deviations, and hence the algorithm is 
successful in recovering the true values of the parameters. Inefficiency 
factors are low, which indicate that the MCMC draws mix well and the 
proposed algorithm is efficient. This pattern is observed at all quantiles 
and across simulation studies. For the scale parameter $\sigma$, the 
posterior estimates vary with quantiles, have small standard deviations, and 
low inefficiency factors. The posterior estimates of the components of 
covariance matrix (i.e., $\Omega_{11}, \Omega_{22}$, and $\Omega_{12}$) are 
close to the true values used to generate the data. However, inefficiency 
factors for these parameters tend to be higher as compared to other model 
parameters indicating slower mixing. But, this is not unusual and have 
earlier been reported in the literature \citep{Chib-Jeliazkov-2006}. Such 
slow mixing occurs because $\Omega$ is at second level of modeling hierarchy 
and depends on the data only via $\{ \alpha_{i}\}$, the random effects 
parameter. 

The posterior estimates of the shape parameter $\gamma$ suggest that the
GAL distribution allows considerable flexibility in skewness while modeling 
the data relative to the AL distribution. For example, in Simulation Study~9, 
the posterior mean of $\gamma$
at the 25th (75th) quantile is 1.20 ($-1.28$) which corresponds to a skewness
of $-0.06$ ($0.13$). This is in contrast to the fixed skewness of
$1.64$ and $-1.64$ in the REQ model. Additionally, the posterior mean at
the 50th quantile is $-0.00$ with a high standard deviation ($0.02$), which 
implies zero skewness. In summary, all estimates of $\gamma$ suggest a 
skewness that is much closer to the skewness of underlying data 
generating process.

%------------------------------- Table 2 -------------------------------------
\begin{table}[!t]
\centering \footnotesize \setlength{\tabcolsep}{4pt}
\setlength{\extrarowheight}{1.25pt}
\setlength\arrayrulewidth{1pt} \caption{Quantile log marginal likelihoods for 
the FREQ and REQ models in 9 simulations studies.}\vspace{10pt}
\begin{tabular}{llr rrr rr}
\toprule
&& \textsc{10th qtl} & \textsc{25th qtl}& \textsc{50th qtl} &\textsc{75th 
qtl} & \textsc{90th qtl} & \\
\midrule		
%-----------------------------------------------------------------------------
\textsc{SS1-FREQ} &&  $-7.680$ & $-6.399$  & $-6.920$  & $-6.689$ &  
$-7.721$  & \vspace{0.1cm}  \\
\textsc{SS1-REQ}  && $-18.120$ & $-8.086$  & $-6.366$  & $-6.985$ & 
$-14.747$  &  \\
\midrule		
%-----------------------------------------------------------------------------
\textsc{SS2-FREQ} && $-11.168$ & $-7.970$  & $-9.759$  & $-8.382$ &  
$-10.687$  & \vspace{0.1cm}  \\
\textsc{SS2-REQ}  && $-23.991$ &$-14.136$  & $-8.760$  &$-12.432$ & 
$-23.513$  &\\
\midrule		
%-----------------------------------------------------------------------------
\textsc{SS3-FREQ} && $-11.579$ & $-9.945$  &$-11.208$  & $-9.930$ &  
$-11.764$  & \vspace{0.1cm}  \\
\textsc{SS3-REQ}  && $-25.833$ &$-13.133$  & $-9.777$  &$-14.018$ & 
$-25.005$  &\\
\midrule
%-----------------------------------------------------------------------------
%-----------------------------------------------------------------------------
\textsc{SS4-FREQ} &&  $-9.220$ & $-7.579$  & $-8.693$  & $-8.147$ &  
$-10.228$  & \vspace{0.1cm}  \\
\textsc{SS4-REQ}  && $-18.419$ & $-8.695$  & $-7.690$  &$-10.648$ & 
$-20.058$  &\\
\midrule
%-----------------------------------------------------------------------------
\textsc{SS5-FREQ} && $-12.589$ &$-10.454$  &$-12.521$  &$-10.894$ &  
$-12.415$  & \vspace{0.1cm}  \\
\textsc{SS5-REQ}  && $-22.161$ &$-13.345$  &$-11.037$  &$-12.783$ & 
$-21.952$  &\\
\midrule
%-----------------------------------------------------------------------------
\textsc{SS6-FREQ} && $-15.598$ &$-12.799$  &$-14.420$  &$-12.956$ & 
$-15.902$  & \vspace{0.1cm}  \\
\textsc{SS6-REQ}  && $-24.590$ &$-15.058$  &$-12.707$  &$-15.391$ & 
$-25.218$  &\\
\midrule
%-----------------------------------------------------------------------------
%-----------------------------------------------------------------------------
\textsc{SS7-FREQ} && $-12.186$ &$-10.126$  &$-10.565$  & $-9.792$ & $ 
-11.709$  & \vspace{0.1cm}  \\
\textsc{SS7-REQ}  && $-20.430$ &$-12.026$  & $-9.363$  &$-10.996$ & 
$-18.694$  &\\
\midrule
%-----------------------------------------------------------------------------
\textsc{SS8-FREQ} && $-13.902$ &$-12.210$  &$-13.428$  &$-12.227$ & 
$-14.045$  & \vspace{0.1cm}  \\
\textsc{SS8-REQ}  && $-21.790$ &$-13.851$  &$-11.724$  &$-13.577$ & 
$-21.250$  &\\
\midrule
%-----------------------------------------------------------------------------
\textsc{SS9-FREQ} && $-15.463$ &$-14.450$  &$-15.694$  &$-14.454$ & 
$-16.229$  & \vspace{0.1cm}  \\
\textsc{SS9-REQ}  && $-22.591$ &$-15.066$  &$-13.592$  &$-15.172$ & 
$-23.318$  &\\
%-----------------------------------------------------------------------------
\bottomrule
\end{tabular}
\label{Table:SimModelfit}
\end{table}
%-----------------------------------------------------------------------------

The above discussion clearly establishes the flexibility of FREQ model 
\emph{vis-a-vis} the REQ model. However, a more pertinent question is what do 
we gain from this flexibility? In particular, does the flexibility translate 
to a better model fit that may justify the additional effort required to 
estimate the FREQ model? To answer this question, we also estimate the REQ 
model using Algorithm~\ref{alg:algorithm2}, compute the log-marginal 
likelihoods for the two models, and report them in 
Table~\ref{Table:SimModelfit}. 
Looking at the results from Simulation Study~9, we see that the log-marginal 
likelihood is higher for the FREQ model at the 10th, 25th, 75th, and 90th 
quantiles. This lends evidence that the FREQ model is better supported by the 
data. Moreover, the difference between log-marginal likelihoods of the FREQ 
and REQ models is higher at the 10th or 90th quantiles as compared to 25th or 
75th quantiles, which suggests that the gains from flexibility increase as we 
move towards the tail of the distribution. At 50th quantile, the log-marginal 
likelihood is higher for the REQ framework, but here a direct comparison of 
marginal likelihoods may be misleading. This is because the posterior mean of 
$\gamma$ is statistically equivalent to zero, thus pointing to a zero 
skewness framework, i.e., the REQ model. Therefore, both frameworks are 
rather equivalent at the 50th quantile. Across the simulation studies, we 
observe that the FREQ framework better fits the data at non-50th quantiles. 
Whereas, for the 50th quantile, the two models are 
equivalent. In conclusion, the FREQ model tends to be a better model than its 
REQ counterpart at quantiles away from the median. The appropriate model (AL 
vs. GAL) will be application- and data-specific. However, in order for a 
researcher to have a more thorough understanding of the data distribution, 
both approaches should be considered and compared in a model comparison 
analysis. We conduct this exercise in the next section where we study 
residential rental rates in the US.

%-------------------------------------------------------------------------------
\section{Application}\label{sec:application}
%------------------------------------------------------------------------------

The US housing sector has received considerable attention as a result of the 
Global Financial Crisis (GFC) when mortgage delinquencies and foreclosures 
increased. Homeownership rates fell from 69\% in 2004 to 62\% in 2016. 
Because of the nexus between house prices and residential rents (see 
\cite{Loewenstein-Willen} for a recent study), a drop in homeownership is 
likely to increase demand for rental units. However, the GFC also featured a 
drop in household income and an increase in unemployment.\footnote{Aggregate 
figures for mortgage delinquencies, unemployment, household income, and 
homeownership are available on FRED.} The concurrence of these events, 
theoretically, leads to ambiguous effects on rental rates. In this 
application, we take an empirical approach and explore changes to residential 
rental rates in the post-GFC United States. In particular, we examine how 
median rental prices in 14,533 zip codes are influenced by unemployment rates 
and mortgage policies. 

%--------------------------  Figure 3 ----------------------------------------
\begin{figure*}[!t]
	\centerline{
		\mbox{\includegraphics[width=7.0in, height=4.5in]{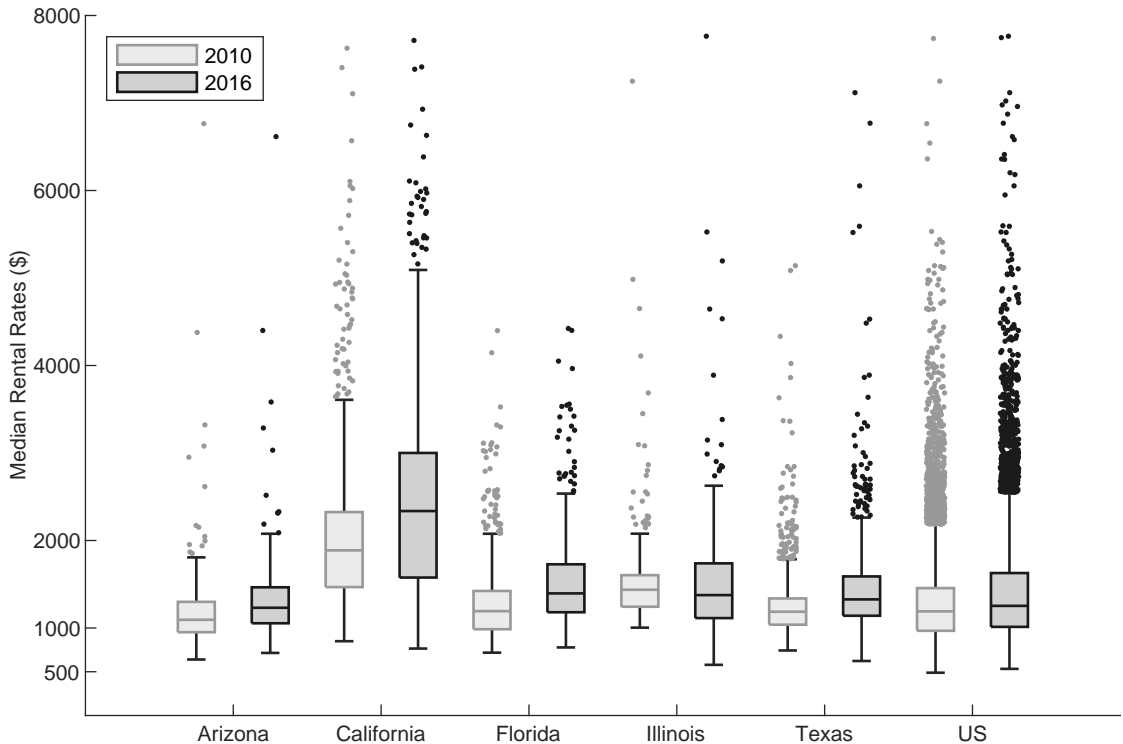}} }
	\caption{Box plots of median rental rates in 5 states and the entire US 
	for 2010 and 2016. Note that the y-axis has been capped at \$8000 for 
	better visual representation.} \label{fig:Boxplot}
\end{figure*}
%-------------------------------------------------------------------------------

In studying rental rates, two concerns must be addressed: (1) heterogeneity because regions of the US vary greatly by housing supply, income, and other factors that influence prices, and (2) skewness in the distribution of prices because that distribution has a large right tail and significant outliers. Exhibiting these concerns, Figure~\ref{fig:Boxplot} presents box plots of median rental rates in 5 states and the entire US for 2010 and 2016. Apparent from the figure are the dramatic differences across the states, where California's lower quartile is above most other states' upper quartile. Additionally, all states display a strong upper (right) skew. Thus, we employ our FREQ model to accommodate heterogeneity and allow for skewness flexibility in the error term. The performance of the FREQ model is tested relative to the REQ model using our novel marginal likelihood approach. This exercise provides important insights for understanding how the data support the different specifications across various quantiles.

%\begin{figure}[tbh]
%\vspace{0.1in}
%\par
%\begin{center}
%\ \psfig{figure=rrdist.eps,width = 0.65\textwidth
%}
%\end{center}
%\par
%\vspace{-0.25in} \caption{A histogram of median residential rental monthly
%prices in 2016. The top panel presents the entire distribution. The bottom
%left panel zooms in on the left tail (quantile 0.10), zip codes with median
%monthly rents less than \$850. The bottom right panel zooms in on the right
%tail (quantile 0.90), zip codes with median monthly rents greater
%than \$2000. } \label{fig:Distribution}
%\end{figure}

There is a vast literature on house prices and rental rates both before and after the Global Financial Crisis. Studies have examined various price determinants including zoning, regulation, and housing supply \citep{Glaeser-etal-2005, Jackson2018}, income differences \citep{Quigley-Raphael-2004}, and tax policy \citep{Chatterjee-Eyignungor-2015}. Many of these studies highlight how heterogeneity in city-specific features, such as average income and land availability, can lead to discrepancies in the effect of the boom and bust on prices. Additionally, a few international papers have focused on the distribution or quantiles of rental prices \citep{Thomschke-2015, Marz-etal-2016, Waltl2018}. We contribute to this literature by implementing a novel quantile regression approach to study rental rates in the United States. Moving beyond mean regression and exploiting large differences in population, income, and economic activity provides a deeper understanding of the determinants of rental markets. Further, our new methodology and model comparison approaches allow us to uncover potential biases that may result from ignored heterogeneity and erroneous distributional assumptions.

%https://www.realpage.com/analytics/distribution-of-apartment-rental-rates/

%-------------------------------------------------------------------------------
\subsection{Data}\label{sec:data}
%------------------------------------------------------------------------------

We construct a novel zip-code-level data set where our outcome variable of interest, $y_{it}$, is the median monthly rental price of zip code $i$ at year $t$. The sample includes $n=14,533$ zip codes in the United States from 2010-2016 ($T=7$). The residential rental price data come from the Zillow Rental Index (ZRENT). Our covariates include annual controls for each zip code's population, demographics, socioeconomic status, agriculture, property ownership, mortgage characteristics, and unemployment. The covariates are constructed from the Statistics of Income (SOI) Tax Stats, Individual Income Tax Statistics, provided by the Internal Revenue Service (IRS). Table~\ref{Table:DataSummary} presents descriptions and summary statistics of our variables. We proxy for ``population'' using the total number of tax returns filed in the zip code and the remaining variables are generally a function of that measure.

%---------------------------  Table 3 ----------------------------------------
\begin{table}[!t]
\centering \footnotesize \setlength{\tabcolsep}{6pt} \setlength{\extrarowheight}{1.5pt}
\setlength\arrayrulewidth{1pt}
\caption{Data summary.}
\begin{tabular}{lp{7.5cm}r r r r }
\toprule
\textsc{variable}       & Description   & Mean   & SD  &  Max  &  Min \\
\midrule
%------------------------------------------------------------------------------
LnRent          & Logarithm of median monthly rental price       
                &  7.177  &  0.378   &  9.745   & 6.082   \\
SSBfrac         & Fraction of the population receiving social security 
benefits
                &  0.139  &  0.059   &  0.814   & 0       \\
Farmfrac        & Fraction of the population receiving farming credits
                &  0.019  &  0.033   &  0.315   & 0       \\
%------------------------------------------------------------------------------
REfrac          & Fraction of the population with real estate taxes
                &  0.273  &  0.134   &  0.821   & 0       \\
HMRate          & Fraction of the population with home mortgage deductions
                &  0.236  &  0.116   &  0.786   & 0       \\
AltMinRate      & Fraction of the population paying alternative minimum taxes
                &  0.027  &  0.050   &  0.444   & 0       \\
%------------------------------------------------------------------------------
EnergyRate      & Fraction of the population receiving energy tax credits
                &  0.024  &  0.019   &  0.147   & 0       \\
EITCRate        & Fraction of the population receiving earned income tax 
credits         &  0.184  &  0.099   &  0.711   & 0       \\
UnempRate       & Fraction of the population receiving unemployment 
compensation    &  0.072  &  0.043   &  0.518   & 0.001   \\
%------------------------------------------------------------------------------
lAvgAGI         & ln-average adjusted gross income
                &  4.040  &  0.458   &  7.899   & 1.610   \\
lreturn         & ln-number of returns filed (proxy for population) 
                &  8.491  &  1.103   & 10.902   & 4.605   \\
%------------------------------------------------------------------------------
\bottomrule
\end{tabular}
\label{Table:DataSummary}
\end{table}
%------------------------------------------------------------------------------

We model the data using the FREQ and REQ models, where $y_{it}= LnRent_{it}$, $x_{it}$ includes the remaining variables in Table~\ref{Table:DataSummary}, and $z_{it}$ is a constant to control for zip-code-level heterogeneity. Heterogeneity is an important concern. While a researcher can control for a host of demographic, socioeconomic, and location characteristics, much is left unobserved. City-level policies, nearby neighborhood spillovers, and commuting effects may enter the error term, which heavily influence rental prices. Thus, our specifications for both models include zip code random effects. Additionally, we include time dummies to capture aggregate changes to prices.

%-------------------------------------------------------------------------------
\subsection{Training Sample Priors}\label{sec:TSprior}
%------------------------------------------------------------------------------

Prior distributions play an important role in Bayesian inference, particularly in model comparison where marginal likelihoods and hence their ratios, the Bayes factors, become arbitrary with improper priors, or sensitive to the prior with formally proper, but increasingly diffuse priors. Therefore, we employ a training sample approach where we take 10\% of our data as a training sample and retain the remainder as a comparison sample. The data in the training sample are used to construct a first-stage posterior distribution which is used as a proper informative training sample prior when analyzing the comparison sample. Information from the training sample is not lost, as it is now part of the prior density used in evaluating the marginal likelihood over the remaining 90\% of the data.\footnote{When estimating the model on the training sample, we used the following relatively uninformative priors: $\beta \sim N(0_{k}, 25 I_{k})$, $\varphi^{2} \sim IG(12/2, 10/2)$, $\sigma \sim IG(10/2,8/2)$ and $\gamma \sim \mathrm{Unif}(L,U)$, where $(L,U)$ are obtained as mentioned in Section~\ref{sec:meth}.}

%-------------------------------------------------------------------------------
\subsection{Results}\label{sec:results}
%------------------------------------------------------------------------------

%Our model results for US sample are based on 5,000 MCMC iterations after a burn-in of 2,000. The MCMC iterations are lower due to the large size of the dataset, leading to computational burdens while estimating the FREQ model for each quantile (estimation of each quantile takes about 18 hours). In contrast, the results for Arizona, California, and Illinois are based on 10,000 MCMC iterations after a burn-in of 2,500 iterations. Due to the smaller size of the state data, the algorithms for the FREQ model are quick (approximately 20 minutes for Arizona, 100 minutes for California, and 40 minutes for Illinois). The run time for the REQ model is approximately one-fifth of its FREQ counterpart.

Before getting to the parameter estimates, we bring attention to the marginal likelihood results. We compare the performance of the FREQ model, relative to the REQ model, for the full sample of US zip codes, as well as several state-specific models (Arizona, California, and Illinois). We consider these smaller samples of states to empirically explore the model fit when $n$ is smaller and when there are varying degrees of heterogeneity and skewness in the sample. Table~\ref{Table:RentalModelfit} presents the log-marginal likelihood estimates for the four samples across five quantiles. We find that the FREQ model has a higher marginal likelihood than the REQ model in all samples at the 10th, 25th, 75th, and 90th quantiles. The differences, particularly further in the tails, are quite dramatic, giving the FREQ model a posterior model probability of $\approx 1$ over the REQ model. At the 50th quantile, we find that the REQ is the favored model in all samples except California. In fact, at the 50th quantile, $\gamma$ is statistically equivalent to 0 in the FREQ model (for all states except California), implying the AL parameterization is more appropriate. Overall, these results demonstrate strong support from the data for the additional flexibility of the FREQ model. Researchers especially interested in the tail of their distribution of interest should employ this more flexible approach in their applied work to improve model fit.

%------------------------------- Table 4 -------------------------------------

\begin{table}[!b]
\centering \footnotesize \setlength{\tabcolsep}{4pt}
\setlength{\extrarowheight}{1.25pt}
\setlength\arrayrulewidth{1pt} \caption{Logarithm of marginal likelihood across 5 quantiles (10th, 25th, 50th, 75th, and 90th) within the FREQ and REQ framework for  Arizona, California, Illinois, and the entire US.}\vspace{10pt}
\begin{tabular}{llr rrr rr}
\toprule
&& \textsc{10th qtl} & \textsc{25th qtl}& \textsc{50th qtl} &\textsc{75th 
qtl} & \textsc{90th qtl} & \\
\midrule		
%-----------------------------------------------------------------------------
Arizona$-$FREQ 
&&  $-371.53$   & $-318.93$   & $-305.80$  & $-322.42$ &  $-388.23$  & 
\vspace{0.1cm}  \\
Arizona$-$REQ  
&&  $-540.59$   & $-350.09$   & $-301.39$  & $-356.09$ &  $-543.64$  & \\
\midrule		
%-----------------------------------------------------------------------------
California$-$FREQ 
&&  $-394.32$   & $-333.22$   & $-316.31$  & $-323.32$ &  $-368.53$  & 
\vspace{0.1cm}  \\
California$-$REQ  
&&  $-633.98$   & $-387.15$   & $-327.43$  & $-380.68$ &  $-610.52$  & \\
\midrule		
%-----------------------------------------------------------------------------
Illinois$-$FREQ 
&&  $-357.94$   & $-305.63$   & $-289.33$  & $-303.70$ &  $-355.26$  & 
\vspace{0.1cm}  \\
Illinois$-$REQ  
&& $-535.54$    & $-337.10$   & $-284.03$  & $-339.78$ &  $-550.31$  &\\
\midrule
%-----------------------------------------------------------------------------
%-----------------------------------------------------------------------------
US$-$FREQ 
&&  $-398.24$ & $-350.69$  & $-333.37$  & $-345.48$ &  $-392.04$  & 
\vspace{0.1cm}  \\
US$-$REQ  
&& $-600.87$ & $-379.30$  & $-328.52$  & $-378.16$ &  $-609.34$  &\\
%-----------------------------------------------------------------------------
\bottomrule
\end{tabular}
\label{Table:RentalModelfit}
\end{table}
%-----------------------------------------------------------------------------

Turning attention to the parameter estimates, Table~\ref{Table:ResultGAL} presents the results for the FREQ model and Table~\ref{Table:ResultAL} presents the results for the REQ model. We find that unemployment is positively associated with residential rental rates. All else equal, a 1 percentage point increase in the fraction of the population that receives unemployment compensation is associated with an increase in monthly residential rental rates of 0.33\% (at the 50th quantile). In a 2012 speech to the National Association of Home Builders, Ben Bernanke, then Chairman of the Federal Reserve, stated that ``High unemployment and uncertain job prospects may have reduced the willingness of some households to commit to homeownership.'' By not committing to homeownership, individuals shift their preferences toward renting. The increase in demand for rental units puts upward pressure on prices, explaining the positive result. We find that the effect gets incrementally larger at higher quantiles, i.e., regions of the US that are more expensive. As an economy recovers from a crisis, attention should be paid to the price of rental units. Policymakers may want to focus on limiting the upward pressure on these prices as individuals and families may have a more difficult time recovering from the economic downturn if rents increase.

%------------------------------- Table 5 -------------------------------------
\begin{table}[!b]
\centering \footnotesize \setlength{\tabcolsep}{4pt} \setlength{\extrarowheight}{2pt}
\setlength\arrayrulewidth{1pt}
\caption{Results for the entire US data assuming a GAL error distribution --
Posterior mean (\textsc{mean}) and posterior standard deviation (\textsc{sd})
of the parameters.}
\begin{tabular}{l rrr rrr rrr rrr rrr   }
\toprule
%------------------------------------------------------------------------------
& & \multicolumn{2}{c}{10th quantile}  & & \multicolumn{2}{c}{25th quantile}
& & \multicolumn{2}{c}{50th quantile}  & & \multicolumn{2}{c}{75th quantile}
& & \multicolumn{2}{c}{90th quantile} \\
\cmidrule{3-4} \cmidrule{6-7}  \cmidrule{9-10} \cmidrule{12-13} \cmidrule{15-16}
%------------------------------------------------------------------------------
          & &  \textsc{mean} & \textsc{sd}
          & &  \textsc{mean} & \textsc{sd}
          & &  \textsc{mean} & \textsc{sd}
          & &  \textsc{mean} & \textsc{sd}
          & &  \textsc{mean} & \textsc{sd} \\
\midrule
%------------------------------------------------------------------------------
Intercept       && $ 5.64$  & $0.02$  && $ 5.64$  & $0.02$  && $ 5.66$ & $0.02$
                && $ 5.70$  & $0.02$  && $ 5.77$  & $0.02$  \\
%------------------------------------------------------------------------------
SSBfrac         && $-0.89$  & $0.02$  && $ -0.94$  & $0.02$  && $-0.94$ & $0.02$
                && $-0.95$  & $0.02$  && $-0.96$  & $0.02$  \\
%------------------------------------------------------------------------------
Farmfrac        && $ -0.42$  & $0.05$  && $ -0.40$  & $0.05$  && $ -0.39$ & $0.04$
                && $ -0.39$  & $0.05$  && $ -0.42$  & $0.05$  \\
%------------------------------------------------------------------------------
REfrac          && $  0.54$  & $0.03$  && $  0.51$  & $0.03$  && $  0.50$ & $0.03$
                && $  0.50$  & $0.03$  && $  0.51$  & $0.03$  \\
%------------------------------------------------------------------------------
HMrate          && $ -0.30$  & $0.04$  && $ -0.31$  & $0.04$  && $ -0.28$ & $0.03$
                && $ -0.30$  & $0.03$  && $ -0.31$  & $0.04$  \\
%------------------------------------------------------------------------------
AltMinRate      && $ 2.10$  & $0.04$  && $ 2.07$  & $0.04$  && $ 2.02$ & $0.04$
                && $ 2.05$  & $0.04$  && $ 2.09$  & $0.04$  \\
%------------------------------------------------------------------------------
EnergyRate      && $  0.35$  & $0.03$  && $  0.40$  & $0.02$  && $  0.40$ & $0.02$
                && $  0.42$  & $0.02$  && $  0.41$  & $0.03$  \\
%------------------------------------------------------------------------------
EITCrate        && $ -0.61$  & $0.02$  && $ -0.64$  & $0.02$  && $ -0.64$ & $0.02$
                && $ -0.65$  & $0.02$  && $ -0.67$  & $0.02$  \\
%------------------------------------------------------------------------------
UnempRate       && $  0.29$  & $0.02$  && $  0.33$  & $0.02$  && $  0.33$ & $0.01$
                && $  0.33$  & $0.01$  && $  0.34$  & $0.02$  \\
%------------------------------------------------------------------------------
lAvgAGI         && $  0.22$  & $0.00$  && $  0.24$  & $0.00$  && $  0.24$ & $0.00$
                && $  0.25$  & $0.00$  && $  0.25$  & $0.00$  \\
%------------------------------------------------------------------------------
lreturn         && $  0.07$  & $0.00$  && $  0.07$  & $0.00$  && $  0.07$ & $0.00$
                && $  0.07$  & $0.00$  && $  0.06$  & $0.00$  \\
%------------------------------------------------------------------------------
y11             && $  0.01$  & $0.00$  && $  0.01$  & $0.00$  && $  0.01$ & $0.00$
                && $  0.00$  & $0.00$  && $ -0.01$  & $0.00$  \\
%------------------------------------------------------------------------------
y12             && $  0.03$  & $0.00$  && $  0.02$  & $0.00$  && $  0.02$ & $0.00$
                && $  0.02$  & $0.00$  && $  0.01$  & $0.00$  \\
%------------------------------------------------------------------------------
y13             && $  0.08$  & $0.00$  && $  0.07$  & $0.00$  && $  0.07$ & $0.00$
                && $  0.06$  & $0.00$  && $  0.05$  & $0.00$  \\
%------------------------------------------------------------------------------
y14             && $  0.10$  & $0.00$  && $  0.09$  & $0.00$  && $  0.09$ & $0.00$
                && $  0.08$  & $0.00$  && $  0.07$  & $0.00$  \\
%------------------------------------------------------------------------------
y15             && $  0.12$  & $0.00$  && $  0.11$  & $0.00$  && $  0.11$ & $0.00$
                && $  0.10$  & $0.00$  && $  0.09$  & $0.00$  \\
%------------------------------------------------------------------------------
y16             && $  0.12$  & $0.00$  && $  0.12$  & $0.00$  && $  0.12$ & $0.00$
                && $  0.12$  & $0.00$  && $  0.11$  & $0.00$  \\
%------------------------------------------------------------------------------
$\sigma$        && $  0.02$  & $0.00$  && $  0.02$  & $0.00$  && $  0.02$ & $0.00$
                && $  0.02$  & $0.00$  && $  0.02$  & $0.00$  \\
%------------------------------------------------------------------------------
$\gamma$        && $  2.28$  & $0.31$  && $  1.03$  & $0.11$  && $ -0.01$ & $0.01$
                && $ -0.87$  & $0.07$  && $ -2.25$  & $0.28$  \\
%------------------------------------------------------------------------------
$\varphi^{2}$   && $ 0.04$  & $0.00$  && $ 0.04$  & $0.00$  && $ 0.04$ & $0.00$
                && $ 0.04$  & $0.00$  && $ 0.04$  & $0.00$  \\
\bottomrule
%%------------------------------------------------------------------------------
% Marginal Likelihood
% %
% && \multicolumn{2}{c}{$126.25$} && \multicolumn{2}{c}{$105.26$}
% && \multicolumn{2}{c}{$105.18$} && \multicolumn{2}{c}{$106.20$}
% && \multicolumn{2}{c}{$111.41$}  \\
\end{tabular}
\label{Table:ResultGAL}
\end{table}
%-------------------------------------------------------------------------------

%------------------------------- Table 6 -------------------------------------
\begin{table}[!h]
\centering \footnotesize \setlength{\tabcolsep}{4pt} \setlength{\extrarowheight}{2pt}
\setlength\arrayrulewidth{1pt}
\caption{Results for the entire US data assuming an AL error distribution --
Posterior mean (\textsc{mean}) and posterior standard deviation (\textsc{sd})
of the parameters.}
\begin{tabular}{l rrr rrr rrr rrr rrr   }
\toprule
%------------------------------------------------------------------------------
& & \multicolumn{2}{c}{10th quantile}  & & \multicolumn{2}{c}{25th quantile}
& & \multicolumn{2}{c}{50th quantile}  & & \multicolumn{2}{c}{75th quantile}
& & \multicolumn{2}{c}{90th quantile} \\
\cmidrule{3-4} \cmidrule{6-7}  \cmidrule{9-10} \cmidrule{12-13} \cmidrule{15-16}
%------------------------------------------------------------------------------
          & &  \textsc{mean} & \textsc{sd}
          & &  \textsc{mean} & \textsc{sd}
          & &  \textsc{mean} & \textsc{sd}
          & &  \textsc{mean} & \textsc{sd}
          & &  \textsc{mean} & \textsc{sd} \\
\midrule
%------------------------------------------------------------------------------
Intercept       && $ 5.70$  & $0.02$  && $ 5.67$  & $0.02$  && $ 5.66$ & $0.02$
                && $ 5.76$  & $0.02$  && $ 5.86$  & $0.02$  \\
%------------------------------------------------------------------------------
SSBfrac         && $-0.75$  & $0.02$  && $-0.83$  & $0.02$  && $-0.94$ & $0.02$
                && $-0.94$  & $0.02$  && $-0.89$  & $0.02$  \\
%------------------------------------------------------------------------------
Farmfrac        && $ -0.43$  & $0.04$  && $ -0.42$  & $0.04$  && $ -0.39$ & $0.04$
                && $ -0.40$  & $0.05$  && $ -0.44$  & $0.05$  \\
%------------------------------------------------------------------------------
REfrac          && $  0.50$  & $0.03$  && $  0.52$  & $0.03$  && $  0.49$ & $0.03$
                && $  0.47$  & $0.03$  && $  0.47$  & $0.03$  \\
%------------------------------------------------------------------------------
HMrate          && $ -0.19$  & $0.03$  && $ -0.24$  & $0.03$  && $ -0.28$ & $0.03$
                && $ -0.25$  & $0.03$  && $ -0.24$  & $0.03$  \\
%------------------------------------------------------------------------------
AltMinRate      && $ 2.10$  & $0.04$  && $ 2.05$  & $0.04$  && $ 2.03$ & $0.04$
                && $ 2.03$  & $0.04$  && $ 1.99$  & $0.04$  \\
%------------------------------------------------------------------------------
EnergyRate      && $  0.17$  & $0.02$  && $  0.29$  & $0.02$  && $  0.40$ & $0.02$
                && $  0.40$  & $0.02$  && $  0.35$  & $0.02$  \\
%------------------------------------------------------------------------------
EITCrate        && $ -0.57$  & $0.02$  && $ -0.59$  & $0.02$  && $ -0.64$ & $0.02$
                && $ -0.68$  & $0.02$  && $ -0.69$  & $0.02$  \\
%------------------------------------------------------------------------------
UnempRate       && $  0.19$  & $0.02$  && $  0.26$  & $0.01$  && $  0.33$ & $0.01$
                && $  0.34$  & $0.01$  && $  0.36$  & $0.02$  \\
%------------------------------------------------------------------------------
lAvgAGI         && $  0.18$  & $0.00$  && $  0.21$  & $0.00$  && $  0.24$ & $0.00$
                && $  0.24$  & $0.00$  && $  0.24$  & $0.00$  \\
%------------------------------------------------------------------------------
lreturn         && $  0.08$  & $0.00$  && $  0.07$  & $0.00$  && $  0.07$ & $0.00$
                && $  0.06$  & $0.00$  && $  0.06$  & $0.00$  \\
%------------------------------------------------------------------------------
y11             && $  0.03$  & $0.00$  && $  0.02$  & $0.00$  && $  0.01$ & $0.00$
                && $ -0.01$  & $0.00$  && $ -0.02$  & $0.00$  \\
%------------------------------------------------------------------------------
y12             && $  0.04$  & $0.00$  && $  0.04$  & $0.00$  && $  0.02$ & $0.00$
                && $  0.01$  & $0.00$  && $ -0.01$  & $0.00$  \\
%------------------------------------------------------------------------------
y13             && $  0.09$  & $0.00$  && $  0.08$  & $0.00$  && $  0.07$ & $0.00$
                && $  0.04$  & $0.00$  && $  0.03$  & $0.00$  \\
%------------------------------------------------------------------------------
y14             && $  0.12$  & $0.00$  && $  0.11$  & $0.00$  && $  0.09$ & $0.00$
                && $  0.07$  & $0.00$  && $  0.05$  & $0.00$  \\
%------------------------------------------------------------------------------
y15             && $  0.13$  & $0.00$  && $  0.12$  & $0.00$  && $  0.11$ & $0.00$
                && $  0.09$  & $0.00$  && $  0.07$  & $0.00$  \\
%------------------------------------------------------------------------------
y16             && $  0.13$  & $0.00$  && $  0.13$  & $0.00$  && $  0.12$ & $0.00$
                && $  0.11$  & $0.00$  && $  0.10$  & $0.00$  \\
%------------------------------------------------------------------------------
$\sigma$        && $  0.01$  & $0.00$  && $  0.02$  & $0.00$  && $  0.02$ & $0.00$
                && $  0.02$  & $0.00$  && $  0.01$  & $0.00$  \\
%------------------------------------------------------------------------------
$\varphi^{2}$   && $ 0.05$  & $0.00$  && $ 0.05$  & $0.00$  && $ 0.04$ & $0.00$
                && $ 0.05$  & $0.00$  && $ 0.05$  & $0.00$  \\
%\midrule
%%------------------------------------------------------------------------------
% Marginal Likelihood
% %
% && \multicolumn{2}{c}{$107.14$} && \multicolumn{2}{c}{$110.14$}
% && \multicolumn{2}{c}{$108.52$} && \multicolumn{2}{c}{$106.86$}
% && \multicolumn{2}{c}{$102.65$}  \\
\bottomrule
%------------------------------------------------------------------------------
\end{tabular}
\label{Table:ResultAL}
\end{table}
%-------------------------------------------------------------------------------

We also find negative effects from our home mortgage variable (HMrate). Thus, the fraction of the population taking home mortgage tax deductions is negatively associated with rental prices. The ability to deduct mortgage interest on individual income taxes makes homeownership more attractive than renting. This result is important in light of the Tax Cuts and Jobs Act (TCJA), which was signed into law in the United States in 2017. The Act lowered the mortgage deduction limit and put a limit on how much an individual can subtract from their taxable income. Our model results suggest that this decrease in home mortgage deductions puts upward pressure on rental prices, a costly unintended consequence. Our results are in line with \cite{HembreDantas2022}, who find that reductions in homeownership subsidies increase rental payments.

The other results in Tables~\ref{Table:ResultGAL} and~\ref{Table:ResultAL} 
largely align with intuition. Specifically, we find that income is positively 
associated with rental rates. Average adjusted gross income has a positive 
effect across the quantiles and the fraction of the population paying 
alternative minimums (i.e., high-income taxpayers) is also positively 
associated with rental rates. Whereas, the fraction of the population 
claiming earned income tax credits (EITC), which represents low-income 
working individuals, is negatively associated with rental rates. 
Additionally, the year indicators, which are relative to 2010, are positive 
and get incrementally larger, capturing aggregate increases in prices.

%------------------------------------------------------------------------------
\subsubsection{Additional Considerations}\label{sec:results-addcons}
%------------------------------------------------------------------------------

In this section, we present the FREQ and REQ results when the sample is restricted to zip codes in Illinois. The model specifications remain the same as before. We chose Illinois (IL) for two reasons: (1) we wish to explore empirical parameter estimates in a smaller sample setting and (2) IL provides extensive variation in land value from expensive metropolitan regions (e.g., Chicago) to rural, farming areas. Table~\ref{Table:IllinoisGAL} presents the FREQ results and Table~\ref{Table:IllinoisAL} presents the REQ results. Importantly, recall from Table~\ref{Table:RentalModelfit} that the data support the FREQ model over the REQ model at all quantiles except the 50th.

%------------------------------- Table 7 -------------------------------------
\begin{table}[!b]
\centering \footnotesize \setlength{\tabcolsep}{4pt} \setlength{\extrarowheight}{2pt}
\setlength\arrayrulewidth{1pt}
\caption{Results for Illinois assuming a GAL error distribution --
Posterior mean (\textsc{mean}) and posterior standard deviation (\textsc{sd})
of the parameters.}
\begin{tabular}{l rrr rrr rrr rrr rrr   }
\toprule
%------------------------------------------------------------------------------
& & \multicolumn{2}{c}{10th quantile}  & & \multicolumn{2}{c}{25th quantile}
& & \multicolumn{2}{c}{50th quantile}  & & \multicolumn{2}{c}{75th quantile}
& & \multicolumn{2}{c}{90th quantile} \\
\cmidrule{3-4} \cmidrule{6-7}  \cmidrule{9-10} \cmidrule{12-13} \cmidrule{15-16}
%------------------------------------------------------------------------------
          & &  \textsc{mean} & \textsc{sd}
          & &  \textsc{mean} & \textsc{sd}
          & &  \textsc{mean} & \textsc{sd}
          & &  \textsc{mean} & \textsc{sd}
          & &  \textsc{mean} & \textsc{sd} \\
\midrule
%------------------------------------------------------------------------------
Intercept       && $  5.47$  & $0.12$  && $  5.44$  & $0.12$  && $  5.42$ & $0.12$
                && $  5.49$  & $0.12$  && $  5.63$  & $0.13$  \\
%------------------------------------------------------------------------------
SSBfrac         && $ -1.37$  & $0.13$  && $ -1.52$  & $0.13$  && $ -1.62$ & $0.13$
                && $ -1.74$  & $0.13$  && $ -1.74$  & $0.13$  \\
%------------------------------------------------------------------------------
Farmfrac        && $  1.06$  & $0.24$  && $  1.27$  & $0.26$  && $  1.43$ & $0.26$
                && $  1.38$  & $0.25$  && $  1.16$  & $0.26$  \\
%------------------------------------------------------------------------------
REfrac          && $  0.64$  & $0.17$  && $  0.75$  & $0.17$  && $  0.84$ & $0.18$
                && $  0.93$  & $0.18$  && $  0.94$  & $0.18$  \\
%------------------------------------------------------------------------------
HMrate          && $ -0.80$  & $0.18$  && $ -0.90$  & $0.18$  && $ -0.97$ & $0.18$
                && $ -1.02$  & $0.19$  && $ -0.94$  & $0.18$  \\
%------------------------------------------------------------------------------
AltMinRate      && $  0.94$  & $0.18$  && $  0.79$  & $0.19$  && $  0.65$ & $0.20$
                && $  0.61$  & $0.20$  && $  0.66$  & $0.20$  \\
%------------------------------------------------------------------------------
EnergyRate      && $  1.09$  & $0.12$  && $  1.04$  & $0.12$  && $  0.96$ & $0.12$
                && $  0.82$  & $0.12$  && $  0.70$  & $0.13$  \\
%------------------------------------------------------------------------------
EITCrate        && $ -0.51$  & $0.10$  && $ -0.51$  & $0.10$  && $ -0.52$ & $0.10$
                && $ -0.52$  & $0.10$  && $ -0.47$  & $0.10$  \\
%------------------------------------------------------------------------------
UnempRate       && $ -0.05$  & $0.09$  && $ -0.04$  & $0.09$  && $ -0.02$ & $0.09$
                && $  0.03$  & $0.09$  && $ -0.01$  & $0.09$  \\
%------------------------------------------------------------------------------
lAvgAGI         && $  0.27$  & $0.02$  && $  0.28$  & $0.02$  && $  0.30$ & $0.02$
                && $  0.31$  & $0.02$  && $  0.30$  & $0.02$  \\
%------------------------------------------------------------------------------
lreturn         && $  0.09$  & $0.01$  && $  0.10$  & $0.01$  && $  0.10$ & $0.01$
                && $  0.09$  & $0.01$  && $  0.08$  & $0.01$  \\
%------------------------------------------------------------------------------
y11             && $  0.00$  & $0.01$  && $ -0.00$  & $0.01$  && $ -0.01$ & $0.01$
                && $ -0.02$  & $0.01$  && $ -0.03$  & $0.01$  \\
%------------------------------------------------------------------------------
y12             && $  0.02$  & $0.01$  && $  0.01$  & $0.01$  && $  0.01$ & $0.01$
                && $ -0.01$  & $0.01$  && $ -0.02$  & $0.01$  \\
%------------------------------------------------------------------------------
y13             && $  0.04$  & $0.01$  && $  0.04$  & $0.01$  && $  0.03$ & $0.01$
                && $  0.01$  & $0.01$  && $ -0.01$  & $0.01$  \\
%------------------------------------------------------------------------------
y14             && $  0.04$  & $0.01$  && $  0.03$  & $0.01$  && $  0.03$ & $0.01$
                && $  0.01$  & $0.01$  && $ -0.01$  & $0.01$  \\
%------------------------------------------------------------------------------
y15             && $  0.03$  & $0.01$  && $  0.03$  & $0.01$  && $  0.02$ & $0.01$
                && $  0.02$  & $0.01$  && $ -0.00$  & $0.01$  \\
%------------------------------------------------------------------------------
y16             && $  0.02$  & $0.01$  && $  0.02$  & $0.01$  && $  0.02$ & $0.01$
                && $  0.02$  & $0.01$  && $  0.00$  & $0.01$  \\
%------------------------------------------------------------------------------
$\sigma$        && $  0.02$  & $0.00$  && $  0.02$  & $0.00$  && $  0.02$ & $0.00$
                && $  0.02$  & $0.00$  && $  0.02$  & $0.00$  \\
%------------------------------------------------------------------------------
$\gamma$        && $  1.72$  & $0.07$  && $  0.54$  & $0.04$  && $ -0.11$ & $0.04$
                && $ -0.85$  & $0.05$  && $ -2.01$  & $0.08$  \\
%------------------------------------------------------------------------------
$\varphi^{2}$   && $  0.04$  & $0.00$  && $  0.04$  & $0.00$  && $  0.04$ & $0.00$
                && $  0.04$  & $0.00$  && $  0.04$  & $0.00$  \\
\bottomrule
%%------------------------------------------------------------------------------
\end{tabular}
\label{Table:IllinoisGAL}
\end{table}
%-------------------------------------------------------------------------------

%------------------------------- Table 8 -------------------------------------
\begin{table}[!ht]
\centering \footnotesize \setlength{\tabcolsep}{4pt} \setlength{\extrarowheight}{2pt}
\setlength\arrayrulewidth{1pt}
\caption{Results for Illinois assuming an AL error distribution --
Posterior mean (\textsc{mean}) and posterior standard deviation (\textsc{sd})
of the parameters.}
\begin{tabular}{l rrr rrr rrr rrr rrr   }
\toprule
%------------------------------------------------------------------------------
& & \multicolumn{2}{c}{10th quantile}  & & \multicolumn{2}{c}{25th quantile}
& & \multicolumn{2}{c}{50th quantile}  & & \multicolumn{2}{c}{75th quantile}
& & \multicolumn{2}{c}{90th quantile} \\
\cmidrule{3-4} \cmidrule{6-7}  \cmidrule{9-10} \cmidrule{12-13} \cmidrule{15-16}
%------------------------------------------------------------------------------
          & &  \textsc{mean} & \textsc{sd}
          & &  \textsc{mean} & \textsc{sd}
          & &  \textsc{mean} & \textsc{sd}
          & &  \textsc{mean} & \textsc{sd}
          & &  \textsc{mean} & \textsc{sd} \\
\midrule
%------------------------------------------------------------------------------
Intercept       && $ 5.52$  & $0.12$  && $ 5.52$  & $0.03$  && $ 5.41$ & $0.12$
                && $ 5.54$  & $0.13$  && $ 5.80$  & $0.13$  \\
%------------------------------------------------------------------------------
SSBfrac         && $-1.08$  & $0.14$  && $-1.35$  & $0.13$  && $-1.67$ & $0.13$
                && $-1.75$  & $0.14$  && $-1.53$  & $0.13$  \\
%------------------------------------------------------------------------------
Farmfrac        && $  0.09$  & $0.22$  && $  1.11$  & $0.24$  && $ 1.48$ & $0.25$
                && $  1.32$  & $0.25$  && $  0.92$  & $0.24$  \\
%------------------------------------------------------------------------------
REfrac          && $  0.48$  & $0.18$  && $  0.60$  & $0.17$  && $  0.89$ & $0.18$
                && $  0.94$  & $0.18$  && $  0.90$  & $0.17$  \\
%------------------------------------------------------------------------------
HMrate          && $ -0.60$  & $0.18$  && $ -0.76$  & $0.18$  && $ -1.00$ & $0.18$
                && $ -0.89$  & $0.19$  && $ -0.73$  & $0.18$  \\
%------------------------------------------------------------------------------
AltMinRate      && $ 1.09$  & $0.19$  && $ 0.96$  & $0.18$  && $ 0.58$ & $0.20$
                && $ 0.47$  & $0.21$  && $ 0.69$  & $0.22$  \\
%------------------------------------------------------------------------------
EnergyRate      && $  1.23$  & $0.13$  && $  1.17$  & $0.11$  && $  0.91$ & $0.12$
                && $  0.61$  & $0.12$  && $  0.40$  & $0.12$  \\
%------------------------------------------------------------------------------
EITCrate        && $ -0.48$  & $0.10$  && $ -0.51$  & $0.10$  && $ -0.52$ & $0.10$
                && $ -0.43$  & $0.10$  && $ -0.23$  & $0.10$  \\
%------------------------------------------------------------------------------
UnempRate       && $ -0.22$  & $0.09$  && $ -0.14$  & $0.09$  && $ -0.01$ & $0.01$
                && $ -0.05$  & $0.09$  && $ -0.09$  & $0.09$  \\
%------------------------------------------------------------------------------
lAvgAGI         && $  0.23$  & $0.02$  && $  0.26$  & $0.02$  && $  0.30$ & $0.02$
                && $  0.31$  & $0.02$  && $  0.29$  & $0.03$  \\
%------------------------------------------------------------------------------
lreturn         && $  0.10$  & $0.01$  && $  0.10$  & $0.01$  && $  0.10$ & $0.01$
                && $  0.08$  & $0.01$  && $  0.06$  & $0.01$  \\
%------------------------------------------------------------------------------
y11             && $  0.02$  & $0.01$  && $  0.01$  & $0.01$  && $ -0.01$ & $0.01$
                && $ -0.04$  & $0.01$  && $ -0.06$  & $0.01$  \\
%------------------------------------------------------------------------------
y12             && $  0.04$  & $0.01$  && $  0.02$  & $0.01$  && $  0.00$ & $0.01$
                && $ -0.03$  & $0.01$  && $ -0.05$  & $0.01$  \\
%------------------------------------------------------------------------------
y13             && $  0.06$  & $0.01$  && $  0.05$  & $0.01$  && $  0.02$ & $0.01$
                && $ -0.01$  & $0.01$  && $ -0.03$  & $0.01$  \\
%------------------------------------------------------------------------------
y14             && $  0.06$  & $0.01$  && $  0.05$  & $0.01$  && $  0.02$ & $0.01$
                && $  0.00$  & $0.01$  && $ -0.03$  & $0.01$  \\
%------------------------------------------------------------------------------
y15             && $  0.05$  & $0.01$  && $  0.04$  & $0.01$  && $  0.02$ & $0.01$
                && $  0.00$  & $0.01$  && $ -0.02$  & $0.01$  \\
%------------------------------------------------------------------------------
y16             && $  0.03$  & $0.01$  && $  0.03$  & $0.01$  && $  0.02$ & $0.01$
                && $  0.01$  & $0.01$  && $ -0.01$  & $0.01$  \\
%------------------------------------------------------------------------------
$\sigma$        && $  0.01$  & $0.00$  && $  0.02$  & $0.00$  && $  0.02$ & $0.00$
                && $  0.02$  & $0.00$  && $  0.01$  & $0.00$  \\
%------------------------------------------------------------------------------
$\varphi^{2}$   && $ 0.05$  & $0.00$  && $ 0.05$  & $0.00$  && $ 0.04$ & $0.00$
                && $ 0.04$  & $0.00$  && $ 0.04$  & $0.00$  \\
%\midrule
%%------------------------------------------------------------------------------
% Marginal Likelihood
% %
% && \multicolumn{2}{c}{$107.14$} && \multicolumn{2}{c}{$110.14$}
% && \multicolumn{2}{c}{$108.52$} && \multicolumn{2}{c}{$106.86$}
% && \multicolumn{2}{c}{$102.65$}  \\
\bottomrule
%------------------------------------------------------------------------------
\end{tabular}
\label{Table:IllinoisAL}
\end{table}
%-------------------------------------------------------------------------------

In looking at the results for the 10th quantile, a major discrepancy between the FREQ and REQ is apparent. The REQ model results suggest that unemployment compensation is negatively associated with residential rental prices. Meaning, in regions of IL that are inexpensive (10th quantile), an increase in the fraction of the population receiving unemployment compensation should decrease rental rates. However, the FREQ model results suggest that the effect of unemployment is not statistically different from zero (i.e., unemployment has no effect on rental prices). In considering the marginal likelihood results (Table~\ref{Table:RentalModelfit}), we know that the posterior model probability of the FREQ model is approximately 1, relative to the REQ model, demonstrating that the data overwhelmingly support the specification with the flexibility in the skewness of the error provided by the GAL distribution. Thus, the results of the FREQ model are validated, whereas those of the REQ model are negated. 

This example demonstrates the dangers of ignoring skewness in the error distribution. Had a researcher or policymaker solely considered a model with the AL distributional assumption (which is commonly done), they would have arrived at an erroneous conclusion about the relationship between unemployment and rental rates. We caution against this approach and instead motivate researchers to use model comparison to uncover the best model and to especially consider the GAL approach at higher and lower quantiles, where the benefits are most dramatic.

%------------------------------------------------------------------------------
\section{Conclusion}\label{sec:conclusion}
%------------------------------------------------------------------------------

This article has considered the Bayesian analysis of a random effects quantile regression model for panel data under the generalized asymmetric Laplace distribution, which eliminates the dependence of distributional skewness on the quantile parameter. New computationally efficient MCMC sampling algorithms have been developed for parameter estimation, as well as model comparison, in both the FREQ and REQ versions of the model. Key to the improved properties of our posterior simulator is the idea of carefully designed parameter blocking. Various features of the proposed modeling framework and estimation methodology have been studied in simulation studies.

The paper has also devoted considerable attention to studying the behavior of 
U.S. residential rental rates following the Global Financial Crisis. Our 
methodology fits this purpose very well due to the strong right skew of 
rental rates and the extensive heterogeneity at the zip-code level across 
different regions. Our results reveal that unemployment has positive effects 
on rental rates and mortgage deductions have negative effects. Regions of the 
U.S. characterized by high unemployment also exhibited declines in 
homeownership, leading to an increase in demand for rental units and putting 
upward pressure on prices. The negative effect of mortgage deductions sheds 
light on the unintended consequences of the Tax Cuts and Jobs Act (TCJA) as a 
potential contributor to the large increases in rental prices since 2017.

Based on our model comparisons, we find that the data overwhelmingly support 
the FREQ model in various subsamples and at all quantiles, especially away 
from the median, suggesting that researchers interested in the tails of the 
distribution could find the more flexible GAL modeling framework decidedly 
more useful.

%------------------------------------------------------------------------------
%-------------------- Beginning of Appendices --------------------------------
%------------------------------------------------------------------------------
\clearpage
\newpage \appendix \renewcommand\thesection{Appendix \Alph{section}.}
%\setcounter{secnumdepth}{-1} % This code removes the numbering of the appendix
%------------------------------------------------------------------------------

%------------------------------------------------------------------------------
\section{\textbf{Conditional Densities in the FREQ 
model}}\label{AppendixA:FREQ}

We utilize the joint posterior density of the FREQ model, given by
Equation~\eqref{eq:jointPosterior}, to derive the conditional posterior
densities of our objects of interest. The principle behind the derivation is
to collect all terms involving the parameter of interest and identifying its
distribution, while holding all other parameters fixed. The derivation of the
conditional posteriors below follows the sequence in
Algorithm~\ref{alg:algorithm1}.

\vspace{0.5pc}
%-------------------------------------------
\textbf{(1)} The parameters $(\beta, \alpha)$ are sampled in a block to
account for possible correlation between the parameters and reduce
autocorrelation in the MCMC draws. The joint posterior of $(\beta, \alpha)$
can be expressed as,
%-------------------------
\begin{equation*}
\begin{split}
\pi(\beta, \alpha |y,\nu,h,\sigma,\gamma, \Omega)
& = \pi(\beta| y,\nu,h,\sigma,\gamma, \Omega)
\pi(\alpha |y,\beta,\nu,h,\sigma,\gamma,\Omega) \\
& = \pi(\beta| y,\nu,h,\sigma,\gamma,\Omega)
\prod_{i=1}^{n}\pi(\alpha_{i} |y,\beta,\nu,h,\sigma,\gamma,\Omega).
\end{split}
\end{equation*}
%-------------------------
We first sample $\beta$ marginally of $\alpha$ and then draw $\alpha$
conditional on $\beta$ and other model parameters.

\text{(a)} To find the conditional posterior density $\pi(\beta|
y,\nu,h,\sigma,\gamma,\Omega)$, we integrate out $(\alpha_{i},u_{i})$
from the model,
%--------------------------
\begin{equation}
y_{i} = X_{i}\beta + Z_{i} \alpha_{i} + A\nu_{i} + C|\gamma|h_{i}
+  \Lambda_{i}^{1/2} u_{i} \notag
\end{equation}
%--------------------------
where $\alpha_{i} \sim N(0_{l}, \Omega_{l \times l})$ and $u_{i} \sim
N(0_{T_{i}}, I_{T_{i}})$. This implies that $y_{i}|\beta, \nu, h, \gamma,
\sigma,\Omega $ follows a normal distribution with mean,
%--------------------------
\begin{equation}
E(y_{i}) = X_{i}\beta + A\nu_{i} + C|\gamma|h_{i} \notag
\end{equation}
and covariance,
%--------------------------
\begin{align}
V_{i} & = E\left[ (y_{i}-E(y_{i}) )(y_{i}-E(y_{i}) )'\right] \notag\\
& = E \left[ (Z_{i}\alpha_{i} + \Lambda_{i}^{1/2} u_{i} )
(Z_{i}\alpha_{i} + \Lambda_{i}^{1/2} U_{i} )'\right] \notag  \\
& = Z_{i}E(\alpha_{i}\alpha')_{i}Z'_{i}   +
\Lambda_{i}^{1/2} E(u_{i} u'_{i}) \Lambda_{i}^{1/2}  \notag  \\
& = Z_{i} \Omega Z'_{i} + \Lambda_{i}, \notag
\end{align}
%--------------------------
i.e., $y_{i}| \beta,\nu,h,\sigma,\gamma,\Omega \sim N(X_{i}\beta +
A\nu_{i} + C|\gamma|h_{i}, \; Z_{i} \Omega Z'_{i} + \Lambda_{i} )$ for 
$i=1,2,\cdots,n$. Thus
it follows that the conditional posterior of $\beta$ can be derived as,
%--------------------------
\begin{align*}
& \pi(\beta|y,\nu,h,\sigma,\gamma,\Omega)
\propto f(y|\beta,\nu,h,\sigma,\gamma,\Omega)
\times \pi(\beta) \\
%---------
& \propto \exp\bigg\{-\frac{1}{2} \bigg[ \sum_{i=1}^{n} (y_{i} - X_{i}\beta -
A\nu_{i}-
C |\gamma| h_{i})'V_{i}^{-1} (y_{i} - X_{i}\beta - A\nu_{i}-  C |\gamma| h_{i})  \\
& \hspace{1in} + (\beta - \beta_{0})' B_{0}^{-1}(\beta - \beta_{0})\bigg]
\bigg\}  \\
%---------
& \propto \exp \bigg\{-\frac{1}{2} \bigg[ \sum_{i=1}^{n}
(y_{i} - A\nu_{i} - C|\gamma|h_{i})'V_{i}^{-1}X_{i}\beta
-\beta' \sum_{i=1}^{n}X'_{i}V_{i}^{-1} ( y_{i} - A\nu_{i} - C|\gamma|h_{i} )  \\
& \hspace{1in} + \beta'  \left(\sum_{i=1}^{n}X'_{i}V_{i}^{-1}X_{i} \right)\beta +
\beta'  B_{0}^{-1} \beta  - \beta' B_{0}^{-1}\beta_{0} - \beta'_{0} B_{0}^{-1}
\beta
\bigg]	\bigg\} \\
%---------
& \propto \exp \bigg\{-\frac{1}{2} \bigg[\beta' \tilde{B}^{-1}\beta -
\beta'\tilde{B}^{-1} \tilde{\beta} - \tilde{\beta}'\tilde{B}^{-1}\beta
+ \tilde{\beta}'\tilde{B}^{-1}\tilde{\beta}
- \tilde{\beta}'\tilde{B}^{-1}\tilde{\beta} \bigg] \bigg\}\\
%---------
& \propto \exp \Big\{-\frac{1}{2} (\beta-\tilde{\beta})'\tilde{B}^{-1}
(\beta-\tilde{\beta})  \Big\},
\end{align*}
%--------------------------
where the posterior precision matrix $\tilde{B}^{-1}$ and the posterior mean
$\tilde{\beta}$ are defined as follows:
%--------------------------
\begin{equation*}
\tilde{B}^{-1} = \bigg(\sum_{i=1}^{n} X'_{i}V_{i}^{-1}X_{i} + B_{0}^{-1} \bigg)
\hspace{.1in} \mathrm{and} \hspace{.1in}
\tilde{\beta} = \tilde{B}\bigg( \sum_{i=1}^{n} X'_{i}V_{i}^{-1}(y_{i}- A\nu_{i} -
C|\gamma| h_{i})  + B_{0}^{-1} \beta_{0} \bigg).
\end{equation*}
%--------------------------
Hence, the conditional posterior is a normal distribution and $\beta|y
,\nu,h,\sigma,\gamma,\Omega \sim N(\tilde{\beta}, \tilde{B})$.

\vspace{0.5pc}

%xxxxxxxxxxxxxxxxxxxxxxxxxxxxxxxxxxxxxxxxxxxxxxxxxxxxxxxxxxxxxxxxxxxxxxxxxxxxxx
\text{(b)} The conditional posterior distribution of $\alpha_{i}$ can be
derived as,
%--------------------------
\begin{align*}
& \pi(\alpha_{i}|y, \beta, \nu, h, \sigma, \gamma, \Omega)
\propto f(y_{i}| \beta, \alpha_{i},\nu, h, \sigma, \gamma,\Omega) \times
\pi(\alpha_{i} | \Omega) \notag \\
%-------
& \propto \exp \bigg\{-\frac{1}{2} \big[ (y_{i} - X_{i} \beta - Z_{i} \alpha_{i} -
A\nu_{i} - C|\gamma|h_{i} )'\Lambda_{i}^{-1}
( y_{i} - X_{i} \beta - Z_{i}\alpha_{i} - A\nu_{i} -
C|\gamma|h_{i}) \big]\\
& \hspace{1in} - \frac{1}{2} \; \alpha'_{i} \Omega^{-1} \alpha_{i} \bigg\}
\notag \\
%-------
& \propto \exp \bigg\{ -\frac{1}{2} \bigg[ (y_{i} - X_{i} \beta - A\nu_{i} -
C|\gamma|h_{i} )' \Lambda_{i}^{-1}Z_{i}\alpha_{i} + \alpha'_{i}
Z'_{i}\Lambda_{i}^{-1}Z_{i}\alpha_{i} \\
& \hspace{.8in}- \alpha'_{i}Z'_{i}\Lambda_{i}^{-1} (y_{i} - X_{i} \beta - A\nu_{i} -
C|\gamma|h_{i} ) + \alpha'_{i} \Omega^{-1} \alpha_{i}	\bigg]	\bigg\}  \\
%-------
& \propto \exp \bigg\{- \frac{1}{2} \Big[ \alpha_{i} \tilde{A}_{i}^{-1} \alpha_{i}
- \alpha'_{i} \tilde{A}_{i}^{-1} \tilde{a_{i}} - \tilde{a}_{i}' \tilde{A}_{i}^{-1}
\alpha_{i} + \tilde{a}_{i}' \tilde{A}_{i}^{-1} \tilde{a}_{i} - \tilde{a}_{i}'
\tilde{A}_{i}^{-1} \tilde{a}_{i} \Big] 	\bigg\} \notag \\
%-------
& \propto \exp \left\{ - \frac{1}{2} (\alpha_{i} - \tilde{a}_{i})'
\tilde{A}_{i}^{-1} (\alpha_{i} - \tilde{a}_{i})	\right\},
\end{align*}
%--------------------------
where the posterior precision $\tilde{A}_{i}^{-1}$ and the posterior mean
$\tilde{a}_{i}$ are as follows:
%--------------------------
\begin{equation*}
\tilde{A_{i}}^{-1}= \bigg( Z'_{i}\Lambda_{i}^{-1}Z_{i} + \Omega^{-1} \bigg)
\hspace{0.15in} \mathrm{and} \hspace{0.15in}
\tilde{a}_{i} = \tilde{A}_{i} \Big( Z'_{i}\Lambda_{i}^{-1} (y_{i} - X_{i}\beta
-A\nu_{i} - C|\gamma|h_{i} ) \Big).
\end{equation*}
%--------------------------	
Hence, the conditional posterior is a normal distribution and $\alpha_{i}|y,
\beta, \nu, h, \sigma, \gamma, \Omega \sim N( \tilde{a}_{i},
\tilde{A}_{i} )$ for $i=1,2,\cdots,n$.

\vspace{0.5pc}

%xxxxxxxxxxxxxxxxxxxxxxxxxxxxxxxxxxxxxxxxxxxxxxxxxxxxxxxxxxxxxxxxxxxxxxxxxxxxxx
\textbf{(2)} The conditional posterior distribution of $\Omega$ is
relatively simple and is derived as shown below,
%--------------------------
\begin{align*}
\pi(\Omega|y, \alpha ) & \propto \prod_{i=1}^{n} \Big[ \pi(\alpha_{i}|
\Omega) \Big] \times \pi(\Omega) \notag \\
%-----------
& \propto \prod_{i=1}^{n} \left[ |\Omega|^{-1/2} \exp \left\{-\frac{1}{2}
\; \alpha'_{i}\Omega^{-1}\alpha_{i}	\right\} \right] \times
|\Omega|^{-(\omega_{0}+l+1)/2} \exp \left\{ -\frac{1}{2} \;
\textrm{tr}(\Omega^{-1} 
O_{0})\right\}  \notag \\
%-----------
& \propto |\Omega|^{-(n+\omega_{0}+l+1)/2} \exp \left\{ -\frac{1}{2} \; 
\textrm{tr} \left[ \Omega^{-1} \left( \sum_{i=1}^{n} \alpha_{i} \alpha'_{i} + 
O_{0} \right) \right] \right\},
\end{align*}
%--------------------------	
which is recognized as the kernel of an inverse-Wishart distribution. Hence,
$\Omega|y, \alpha \sim IW (\tilde{\omega}, \tilde{O})$, where,
$\tilde{\omega} = n + \omega_{0}$ and $\tilde{O} = \sum_{i=1}^{n}
\alpha_{i}\alpha'_{i} + O_{0}$.
%--------------------------	

\vspace{0.5pc}

%xxxxxxxxxxxxxxxxxxxxxxxxxxxxxxxxxxxxxxxxxxxxxxxxxxxxxxxxxxxxxxxxxxxxxxxxxxxxxx
\textbf{(3)} The parameters $(\sigma,\gamma)$ are jointly sampled marginally
of $(\nu,h)$ from the joint posterior which is proportional to the likelihood
$f_{GAL}(y|\beta,\alpha,\sigma,\gamma)$ times the prior distributions
$\pi(\beta, \alpha, \sigma,\gamma)$ given by
Equation~\eqref{eq:fulllikelihood} and Equation~\eqref{eq:priors},
respectively. Collecting terms involving $(\sigma,\gamma)$ do not yield a
tractable distribution, so $(\sigma,\gamma)$ are sampled using a random-walk
MH algorithm. Here, joint sampling increases algorithmic efficiency by
reducing the autocorrelation in the MCMC draws of $(\sigma, \gamma)$. The
proposed draw $(\sigma',\gamma')$ are generated from a bivariate truncated
normal distribution $BTN_{(0, \infty) \times (L,U)} \big((\sigma_{c},
\gamma_{c}), \iota^{2} \hat{D} \big)$, where $(\sigma_{c}, \gamma_{c})$ are
the current values, $\iota$ is the tuning factor and $\hat{D}$ is the
negative inverse of the Hessian obtained by maximizing the logarithm of the
likelihood with respect to $(\sigma,\gamma)$ with
$\beta$ set at the pooled ordinal least squares estimate. We accept
$(\sigma',\gamma')$ with MH probability,
%--------------------------
\begin{equation*}
\alpha_{MH}(\sigma_{c},\gamma_{c}; \sigma',\gamma') = \min \bigg\{0,\ln\bigg[
\frac{f(y,\alpha|\beta, \sigma', \gamma') \, \pi(\beta, \alpha,
\sigma', \gamma')} {f(y, \alpha|\beta, \sigma_{c},\gamma_{c}) \,
\pi(\beta, \alpha, \sigma_{c},\gamma_{c} )} \;
\frac{\pi(\sigma_{c}, \gamma_{c} | (\sigma',\gamma'), \iota^{2} \hat{D})}
{\pi(\sigma',\gamma'| (\sigma_{c}, \gamma_{c}),\iota^{2} \hat{D} )}\bigg] \bigg\};
\end{equation*}
%--------------------------
where, $f(y,\alpha|\cdot)$ denotes the full likelihood given by 
Equation~\eqref{eq:fulllikelihood}, $\pi(\beta, \sigma,\gamma )$ 
denotes the prior distributions given in Equation~\eqref{eq:priors}, and 
$\pi(\sigma_{c}, \gamma_{c} | (\sigma',\gamma'), \iota^{2} \hat{D})$ denotes 
the bivariate truncated normal
probability with mean $(\sigma',\gamma')$  and covariance $\iota^{2} \hat{D}$
and \emph{vice-versa}; otherwise, the current value $(\sigma_{c},\gamma_{c})$
is repeated in the next MCMC iteration. Note that the parameters $(A,B,C)$ is
a function of $p$ which in turn is dependent on $p_{0}$ and $\gamma$.
%------------------------------------------------------------------------------

\vspace{0.5pc}

\textbf{(4)} To derive the conditional posterior distribution of $\nu_{it}$,
we need to work element wise as follows:
%--------------------------
\begin{align*}
& \pi(\nu_{it}|y_{it},\beta, \alpha_{i}, h_{it},\sigma,\gamma) \\
%-------
& \propto \nu_{it}^{-\frac{1}{2}} \exp \bigg\{-\frac{1}{2} \bigg[
\frac{(y_{it}-x'_{it}\beta - z'_{it}\alpha_{i} - A\nu_{it} - C|\gamma|
	h_{it})^{2}} {\sigma B \nu_{it}}  \bigg] -\frac{\nu_{it}}{\sigma}
\bigg\} \notag \\
%-------
& \propto \nu_{it}^{-\frac{1}{2}} \exp \bigg\{-\frac{1}{2} \bigg[
\frac{(y_{it}-x'_{it}\beta - z'_{it}\alpha_{i} - C|\gamma|h_{it})^{2}}
{\sigma B } \, \nu_{it}^{-1} + \bigg(\frac{A^{2}}{\sigma B}
+ \frac{2}{\sigma} \bigg) \nu_{it} \bigg] \bigg\} \notag \\
%-------
& \propto \nu_{it}^{-\frac{1}{2}} \exp\bigg\{-\frac{1}{2}
\bigg[\chi \nu_{it}^{-1} + \psi_{\nu_{it}} \nu_{it}  \bigg]  \bigg\},
\end{align*}
%--------------------------
where we have used the following notations,
%--------------------------
\begin{equation*}
\chi = \bigg(\frac{A^{2}}{\sigma B} + \frac{2}{\sigma} \bigg)
\hspace{0.25in} \mathrm{and} \hspace{0.25in}
\psi_{\nu_{it}} = \frac{(y_{it}-x'_{it}\beta - s'_{it}\alpha_{i} -
	C|\gamma|h_{it})^{2}}{\sigma B}.
\end{equation*}
%--------------------------
Therefore, we have $\nu_{it}|y_{it}, \beta, \alpha_{i}, h_{it}, \sigma,
\gamma \sim GIG(\frac{1}{2}, \chi, \psi_{\nu_{it}} )$ for all values of $i$
and $t$.
%------------------------------------------------------------------------------

\vspace{0.5pc}

\textbf{(5)} Similar to $\nu_{it}$, the  conditional posterior of $h_{it}$ is
derived element wise as follows:
%--------------------------
\begin{align*}
& \pi(h_{it}|y_{it},\beta,\nu_{it},\sigma,\gamma) \\
%-----
& \propto \exp \bigg\{ -\frac{1}{2} \bigg[ \frac{(y_{it}-x'_{it}\beta
- z_{it}\alpha_{i} - A\nu_{it}- C|\gamma|h_{it})^{2}} {\sigma B \nu_{it}}
 + \frac{h_{it}^{2}}{\sigma^2}  \bigg]\bigg\} \notag \\
%-----
& \propto \exp \bigg\{-\frac{1}{2} \bigg[ \bigg(\frac{1}{\sigma^2} +
\frac{C^{2} \gamma^{2}}{\sigma B \nu_{it}}  \bigg) h_{it}^{2} -
\frac{2 C|\gamma|(y_{i}-x'_{it}\beta - z'_{it} \alpha_{i} - A\nu_{it})}
{\sigma B \nu_{it}} h_{it}\bigg]  \bigg\}  \notag \\
%-----
& \propto \exp \bigg\{-\frac{1}{2} \bigg[(\sigma_{h_{it}}^{2})^{-1}
h_{it}^{2}  - 2 (\sigma_{h_{it}}^{2})^{-1} \mu_{h_{it}} h_{it}\bigg]
\bigg\} \notag \\
%-----
& \propto \exp \bigg\{ -\frac{1}{2} (\sigma_{h_{it}}^{2})^{-1}
(h_{it} - \mu_{h_{it}})^{2} \bigg\}, \notag
\end{align*}
%--------------------------
where, the second last line utilizes the notations,
%--------------------------
\begin{equation*}
\sigma_{h_{it}}^{2} = \bigg(\frac{1}{\sigma^2} + \frac{C^{2}
\gamma^{2}}{\sigma  B \nu_{it}}  \bigg)^{-1}
\hspace{0.25in} \mathrm{and} \hspace{0.25in}
\mu_{h_{it}} = \sigma_{h_{it}}^{2} \bigg( \frac{ C|\gamma|(y_{it}- x'_{it}
\beta - z'_{it} \alpha_{i} - A\nu_{it})}{\sigma B \nu_{it}}  \bigg).
\end{equation*}
%--------------------------
and the last expression is recognized as the kernel of a half-normal
distribution. Hence, we have $h_{it}| y_{it},\beta,\nu_{it},\sigma,\gamma
\sim N^{+}(\mu_{h_{it}}, \sigma^2_{h_{it}})$ for all values of $i$ and $t$.
%-------------------------------------------

%--------------------------------- Bibliography ------------------------------
\clearpage \pagebreak %\nocite{*}
%\pdfbookmark[1]{References}{unnumbered} % To make References as a bookmark in
%pdf
%\section*{References}
\singlespacing
\bibliography{BibFREQ}
\bibliographystyle{jasa}

%\bibliography{C:/1-ARSHAD-OPTIPLEX/Google-Drive/18-References/BibEconometrics}
%\bibliographystyle{C:/1-ARSHAD-OPTIPLEX/Google-Drive/18-References/jasa}

\end{document}